\documentclass[aps, twocolumn,floatfix,pre,showpacs, superscriptaddress]{revtex4-2}

\usepackage{amssymb,amsmath,amsfonts}
\usepackage{graphicx,graphics}
\usepackage{epsfig}
\usepackage{epstopdf}
\usepackage{amsfonts}
\usepackage{bm,bbm}
\usepackage{amssymb,amsmath,amsfonts}
\usepackage{mathrsfs}
\usepackage{calc}
\usepackage{epsfig}
\usepackage{float}
\usepackage{xcolor}
\usepackage{epstopdf}
\usepackage{bm,amssymb,amsmath}
\usepackage{graphicx,color}

\usepackage{ulem}
\usepackage{xspace}

\newcommand{\T}{T_{\rm eff}}

\usepackage[colorlinks=true, citecolor=teal, urlcolor=purple,linkcolor=purple]{hyperref}

\begin{document}

\title{Consistent thermodynamics reconstructed from transitions\\ between nonequilibrium steady-states}


\author{R\'emi Goerlich}
\email{remigoerlich@tauex.tau.ac.il}
\affiliation{Raymond \& Beverly Sackler School of Chemistry, Tel Aviv University, Tel Aviv 6997801, Israel}
\author{Benjamin Sorkin}
\affiliation{Princeton Center for Theoretical Science, Princeton University, Princeton 08544, New Jersey, USA}
\author{Dima Boriskovsky}
\affiliation{Raymond \& Beverly Sackler School of Chemistry, Tel Aviv University, Tel Aviv 6997801, Israel}
\author{Lu\'is B. Pires}
\affiliation{Departamento de F\'isica, Universidade Federal de Vi\c cosa, CEP 36570-900, Vi\c cosa, Minas Gerais, Brasil}
\author{Benjamin Lindner}
\affiliation{Department of Physics, Humboldt-Universität zu Berlin, Newtonstr 15, 12489 Berlin, Germany}
\affiliation{Bernstein Center for Computational Neuroscience, Haus 2, Philippstr 13, 10115 Berlin, Germany}
\author{Cyriaque Genet}
\email{genet@unistra.fr}
\affiliation{University of Strasbourg and CNRS, CESQ and ISIS, UMR 7006, F-67000 Strasbourg, France}
\author{Yael Roichman}
\email{roichman@tauex.tau.ac.il}
\affiliation{Raymond \& Beverly Sackler School of Chemistry, Tel Aviv University, Tel Aviv 6997801, Israel}
\affiliation{Raymond \& Beverly Sackler School of Physics and Astronomy, Tel Aviv University, Tel Aviv 6997801, Israel}

\begin{abstract}
Constructing a thermodynamic framework for nonequilibrium systems remains a major challenge,
 as quantities such as temperature and free energy often become ambiguous when inferred solely from steady-state properties. 
Here we take a transformation-based approach and experimentally examine transitions between nonequilibrium steady states (NESS). 
 Using an optically trapped microparticle driven by a tunable correlated stochastic force, we generate active-like steady states with controllable noise statistics. By abruptly changing the trap stiffness, we measure the stochastic work, heat, and entropy produced during NESS-to-NESS transformations.
We identify a state-dependent effective temperature that restores the second law for these transitions, enabling the definition of a generalized work that incorporates the consequence of the nonequilibrium fluctuations.
With this quantity, we derive and experimentally verify a Crooks-like fluctuation relation linking work distributions to a nonequilibrium free-energy difference defined through the effective temperature. Finally, we establish a fluctuation–response relation for the positional variance following stiffness changes.
We demonstrate that this relation is key to distinguishing systems that can be described by a unique effective temperature (i.e., those under equilibrium or white-noise conditions) from those under colored-noise, where an equilibrium-like response cannot be restored. These results delineate the applicability and limits of effective-temperature thermodynamics in driven systems.
\end{abstract}

\maketitle

Stochastic thermodynamics relates thermodynamic observables to mesoscopic systems, defining heat, work, and entropy along individual stochastic trajectories \cite{SekimotoBook, Sekimoto1998, Seifert2005, seifert2012}. Within this framework, fluctuation theorems recast the second law in statistical terms \cite{jarzynski1997nonequilibrium, crooks1999entropy, crooks2000path, kurchan1998fluctuation, esposito2010three} and offer practical routes to measuring free-energy differences. Experimental advances in the control of small fluctuating systems have enabled detailed tests of these relations \cite{CilibertoPRX2017, Bechhoefer2020} and clarified how heat, work, and free energy behave under controlled transformations \cite{liphardt2002equilibrium, collin2005verification, koski2013distribution, martinez2015adiabatic}. These developments have also supported efforts in optimal control \cite{baldovin2023control, guery-odelin_shortcuts_2019, guery-odelin_driving_2023, patra_shortcuts_2017, Plata2021, Schmiedl2007, Rosales2020, Pires2023, Raynal2023, olsen2025harnessing} and the design of microscopic heat engines \cite{blickle2012realization, martinez2017colloidal, martinez2016brownian}.

The application of stochastic thermodynamics to nonequilibrium systems, such as active and driven matter \cite{fodor2016far, speck2016stochastic, szamel2019stochastic, flenner2020active, fodor2021active}, presents several challenges. Out of equilibrium, basic quantities such as temperature \cite{maggi2014generalized, puglisi2017temperature}, pressure \cite{solon2015pressure, junot2017active}, or free energy \cite{esposito2011second} are no longer unambiguously defined and therefore require revised formulations. Central relations valid at thermal equilibrium, including the fluctuation–response relation (FRR) \cite{kubo1966fluctuation, agarwal1972fluctuation, marconi2008fluctuation, dieterich2015single, caprini2021fluctuation, baldovin2022many, Lin22} and classical fluctuation theorems (FT) \cite{szamel2023single}, generally fail under standard definitions of thermodynamic quantities and often demand significant modifications \cite{hatano2001steady, esposito2010three, Argun2016}.
To date, the search for a thermodynamic description of systems out of equilibrium has focused mostly on the study of nonequilibrium steady-states (NESS).
Reconciling generalized temperatures with a more consistent thermodynamic framework, including the first and second laws, remains an open problem \cite{fodor2021active}.

Here, we experimentally develop and test such a theoretical framework, focusing on thermodynamic transformations between nonequilibrium steady states (NESSs).
Using an optically trapped colloidal particle driven by tunable colored noise \cite{Goerlich2022}, we generate controlled NESSs and probe their response to abrupt changes in trap stiffness. This approach highlights aspects of nonequilibrium behavior that are not apparent from steady-state measurements alone. We define an effective temperature for this system, which is consistent with the second law of thermodynamics \cite{sorkin2024second}. In contrast to equilibrium, the system's nonequilibrium nature becomes apparent through the temperature becoming a function of the system's state.
We identify a generalized work and free-energy based on this state-dependent effective temperature. This resulting framework characterizes the energy exchanges during the transformations and permits a generalized Crooks FT, which we verify experimentally.
However, the system’s dynamical response retains signatures of correlated nonequilibrium driving. Together, these results outline a transition-based perspective that may help extend thermodynamic reasoning to a broader class of nonequilibrium systems.

\begin{figure*}[t!]
	\centerline{\includegraphics[width=0.95\linewidth]{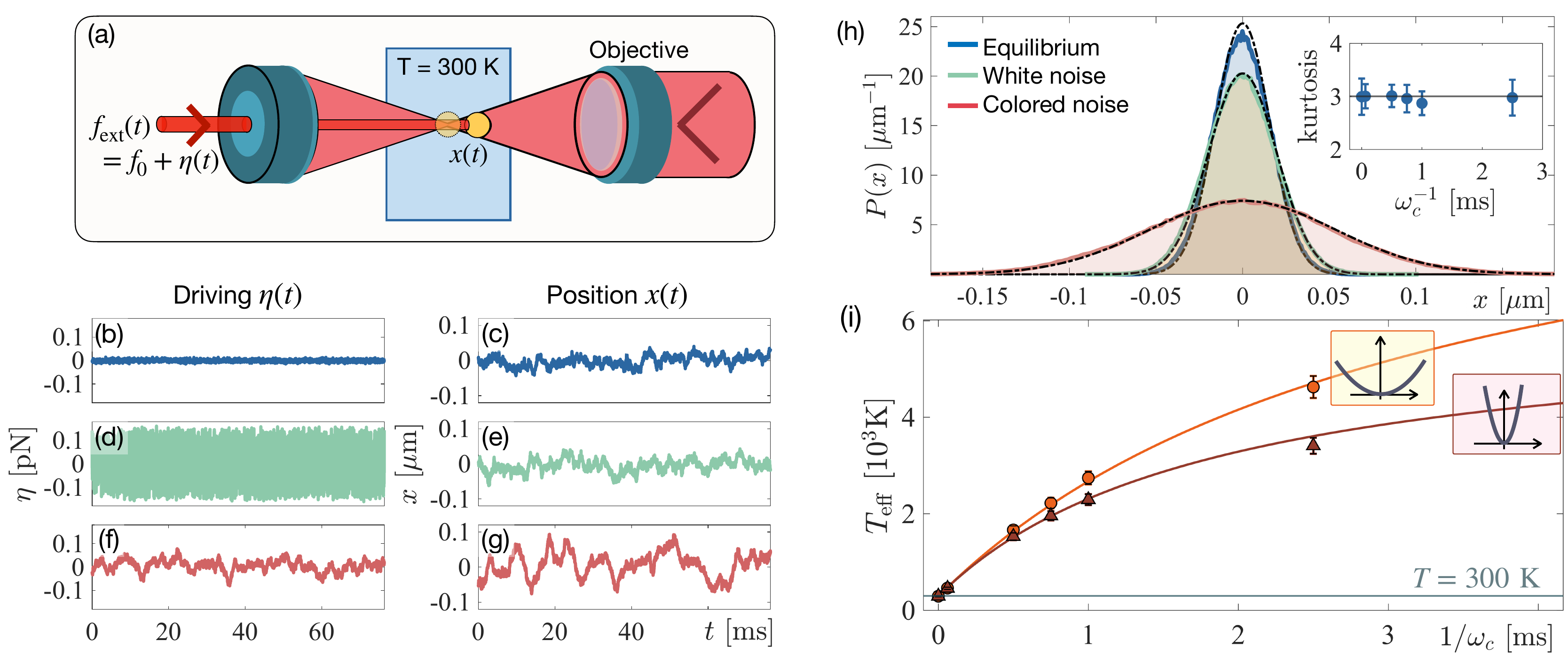}}
	\caption{\textbf{Colloidal particle driven out of equilibrium by colored noise in an optical trap.}
(a) Schematic of the experimental setup (see Appendix).
(b) Vanishing external driving ($\eta = 0$) and (c) corresponding segment of an equilibrium Brownian trajectory $x(t)$ in a harmonic trap.
(d) White-noise driving force (up to cut-off frequency at $2^{14}$ Hz) and (e) the associated trajectory.
(f) Colored-noise driving force and (g) the corresponding trajectory.
(h) Position probability distribution $P(x)$ at equilibrium (blue), under white-noise driving (green), and under colored-noise driving (red), together with the corresponding Gaussian Boltzmann-like distributions (black dashed lines).
Inset: kurtosis $K=\langle x^4\rangle/\langle x^2\rangle^2$ as a function of noise correlation time; $K=3$ indicates Gaussian statistics.
(i) Effective temperature extracted from the Boltzmann factor as a function of the noise correlation time $\omega_{\mathrm{c}}^{-1}$, for initial ($\kappa_{\mathrm{i}}$; orange circles and line) and final ($\kappa_{\mathrm{f}}$; dark red triangles line) trap stiffnesses.
Lines show the analytical prediction [Eq.~(\ref{Eq:Teff})] with experimental parameters.}
	\label{fig:Schema}
\end{figure*}

\section*{State-dependent fluctuations reveal a thermodynamic variable for NESS}
We first examine the stationary fluctuations of a colloidal particle held in an optical trap and driven by a tunable colored stochastic force. Although the dynamics are out of equilibrium, the position distribution remains Gaussian, with a variance that depends systematically on the trap stiffness. This dependence allows us to define a state-dependent effective temperature $T_\text{eff}(k)$, which reduces to the ambient temperature for equilibrium conditions but increases under colored-noise driving. This temperature is not imposed externally but emerges as a measurable property of each nonequilibrium steady state (NESS). Its stiffness dependence is a key signature of the non-Markovian active-like bath.

Specifically, we experimentally realize a driven Brownian diffusion process under a time-dependent external forcing \cite{Goerlich2022}.
In our experiment (Fig. \ref{fig:Schema}(a)), a micron-sized particle is immersed in a cell filled with water at room temperature $T$. It is optically trapped at the waist of a tightly focused laser beam, whose optical potential is well approximated by a quadratic profile, $U(x, \kappa) = \kappa x^2/2$.
The stiffness  $\kappa=\kappa(t)$ of the potential can be dynamically controlled by adjusting the intensity of the laser beam (see details in Appendix).

To generate a nonequilibrium environment, we subject the particle to another external force, $f_{\rm ext}(t)$, generated by radiation pressure from a second laser beam \cite{li2019subfemtonewton, Goerlich2022, Pires2023}. By modulating the intensity of this second beam, the radiation pressure force can be precisely tuned over time.
This second time-dependent external force can be split into two contributions, $f_{\rm ext}(t) = f_0 + \eta(t)$.
The first term, $f_0$, is a constant that simply displaces the equilibrium position of the microsphere in the optical trap, and thus vanishes in the dynamics of the centered process.
The second term, $\eta(t)$, can be an arbitrary time-dependent force with zero mean.
In this work, we program $\eta(t)$ to act as a stochastic forcing, mimicking the effect of an additional bath coupled to the microsphere.

Overall, the dynamics of the tracer's position $x(t)$ is captured by the following overdamped Langevin equation
\begin{equation}
    \dot x(t) = -\frac1\gamma\kappa(t) x(t) + \sqrt{2 D} \xi(t) + \frac{1}{\gamma} \eta(t)
    \label{Eq:Langevin}
\end{equation}
where 
$\gamma$ is the viscous drag coefficient, $D = k_\mathrm{B} T / \gamma$ the thermal diffusion coefficient (with $T$ the surrounding fluid's temperature and $k_\mathrm{B}$ the Boltzmann constant) and $\xi(t)$ a Gaussian white noise with unit variance $\langle \xi(t) \xi(s) \rangle = \delta(t-s)$.

We consider three distinct driving conditions.
In the absence of external noise ($\eta=0$), the particle equilibrates with the surrounding fluid (Fig.~\ref{fig:Schema}(b, c)). When driven by an additional white noise ($\langle \eta(t) \eta(s) \rangle = \sigma^2_\eta \delta(t-s)$), the system behaves as if coupled to an equilibrium bath at an elevated temperature (Fig.~\ref{fig:Schema}(d, e)), a strategy previously used to implement effective thermal reservoirs and mesoscopic heat engines \cite{martinez2013effective, martinez2016brownian, martinez2017colloidal, Pires2023}. Finally, we apply an exponentially correlated Gaussian noise digitally generated by an Ornstein–Uhlenbeck process,
\begin{equation}
    \dot \eta(t) = -\omega_\mathrm{c} \eta(t) + \sqrt{2 \omega_\mathrm{c} \sigma_\eta^2}\theta(t)
    \label{Eq:Noise}
\end{equation}
where $\theta(t)$ is another white noise, $\langle \theta(t) \theta(s) \rangle = \delta(t-s)$ and $\langle \xi(t) \theta(s) \rangle=0$.
The force is therefore characterized by a memory $\langle \eta(t) \eta(s) \rangle = \sigma_\eta^2e^{-\omega_\mathrm{c} |t-s|}$, with correlation time $\omega_\mathrm{c}^{-1}$ and variance $\sigma_\eta^2$. 
With this driving, Eq.~(\ref{Eq:Langevin}) describes the active Ornstein-Uhlenbeck process (AOUP), a well-studied minimal model of active and self-propelled matter \cite{marchetti2013hydrodynamics, maggi2014generalized, elgeti2015physics,Kanazawa2015,Sandford2017, marconi2017heat, fodor2021active, Goerlich2022}. For the latter case (Fig.~\ref{fig:Schema}(f,g)), we will show in the following that the resulting tracer's dynamics differ significantly from equilibrium dynamics \cite{marconi2017heat, Argun2016, Goerlich2022}.
The finite memory displayed by $\eta(t)$ deprives the driven process $x(t)$ of the Markovian property \cite{Sandford2017}, although the two-dimensional process $\{x(t), \eta(t)\}$ is still Markovian.

For constant stiffness $\kappa =  15.1 \pm 2.9 $ pN/$\mu$m and $\eta = 0$, the microsphere equilibrates within the trapping potential, reaching the Boltzmann probability distribution, $P_\mathrm{eq}(x) = [\kappa/(2 \pi k_\mathrm{B} T)]^{1/2} e^{-\kappa x^2 / (2 k_\mathrm{B} T)} $, with a variance obeying the equipartition theorem $\kappa \langle x^2 \rangle = k_\mathrm{B} T$.
Adding either white or colored noise (with correlation time in the range $\omega_\mathrm{c}^{-1}= 0.5 - 2.5 ~ \rm{ms} $; in Fig.~\ref{fig:Schema}, $2.5$ ms) increases the variance, leading to broader steady-state distributions, as already visible in the spreading of the trajectories (Fig.~\ref{fig:Schema}(c, e, g)). Owing to the linearity of the Langevin dynamics Eq.~(\ref{Eq:Langevin}) and the Gaussian nature of the noise, these distributions remain Gaussian (Fig. \ref{fig:Schema}(h)), allowing us to write
\begin{equation}
    P(x, \kappa) = \sqrt{\frac{\kappa}{2 \pi k_\mathrm{B} \T}} e^{-\frac{\kappa x^2}{2 k_\mathrm{B} \T}}
    \label{eq:effboltzmann}
\end{equation}
which defines an effective temperature through the extended equipartition relation $\kappa \langle x^2 \rangle = k_\mathrm{B} \T$. The stiffness $\kappa$ is unambiguously measured in the case where $\eta(t) = 0$.

Unlike equilibrium systems, where $\kappa \langle x^2 \rangle$ remains constant along isotherms, the effective temperature here depends explicitly on the state of the system. For exponentially correlated noise, evaluating the stationary variance yields
\begin{equation}
    \T(\kappa) = T \left( 1 + \frac{\sigma_\eta^2}{\gamma D[\gamma \omega_\mathrm{c} + \kappa]} \right),
    \label{Eq:Teff}
\end{equation}
showing that $\T$ depends not only on the characteristics of the noise, via $\sigma_\eta^2$ and $\omega_\mathrm{c}$, but also on the state of the system, via $\kappa$.
See Appendix for further details.

In Fig.~\ref{fig:Schema}(i), we show $\T(\kappa)$ as a function of the correlation time in two different potentials.
The upper line corresponds to the temperature measured when the system is in a potential with low stiffness $\kappa_\mathrm{i} = 8.51 \pm 1.6$ pN/$\mu$m (circles) while the lower line corresponds to the temperature measured with high stiffness $\kappa_\mathrm{f} = 15.1 \pm 2.9 $ pN/$\mu$m (triangles).
For both stiffnesses, the temperature $\T(\kappa)$ increases with the correlation time of noise, though they do not coincide for a fixed correlation time.

The state-dependent temperature is a direct consequence of the correlated noise, as can be understood from a linear response perspective.
The system, a microsphere under the influence of a potential, possesses a given mechanical susceptibility.
At equilibrium, or when driven by another white noise, all components of the mechanical response function are excited with the same amplitude.
On the other hand, correlated noise with a non-uniform spectrum excites preferentially low-frequency modes of the system \cite{turlier2016equilibrium}.
For different stiffnesses $\kappa$, the distinct response functions changes the coupling to the same correlated bath.
As a consequence, the amplitude of the microsphere's displacement is modified, resulting in an stiffness-dependent effective temperature $\T(\kappa)$. Despite this nonequilibrium character, the harmonic confinement preserves Gaussian steady-state statistics, enabling a thermodynamic description in which $\T(\kappa)$ plays a central role in quantifying energy exchanges.

\section*{NESS-to-NESS transformation reconstruct a second law}

\begin{figure}[t!]
    \centerline{\includegraphics[width=1\linewidth]{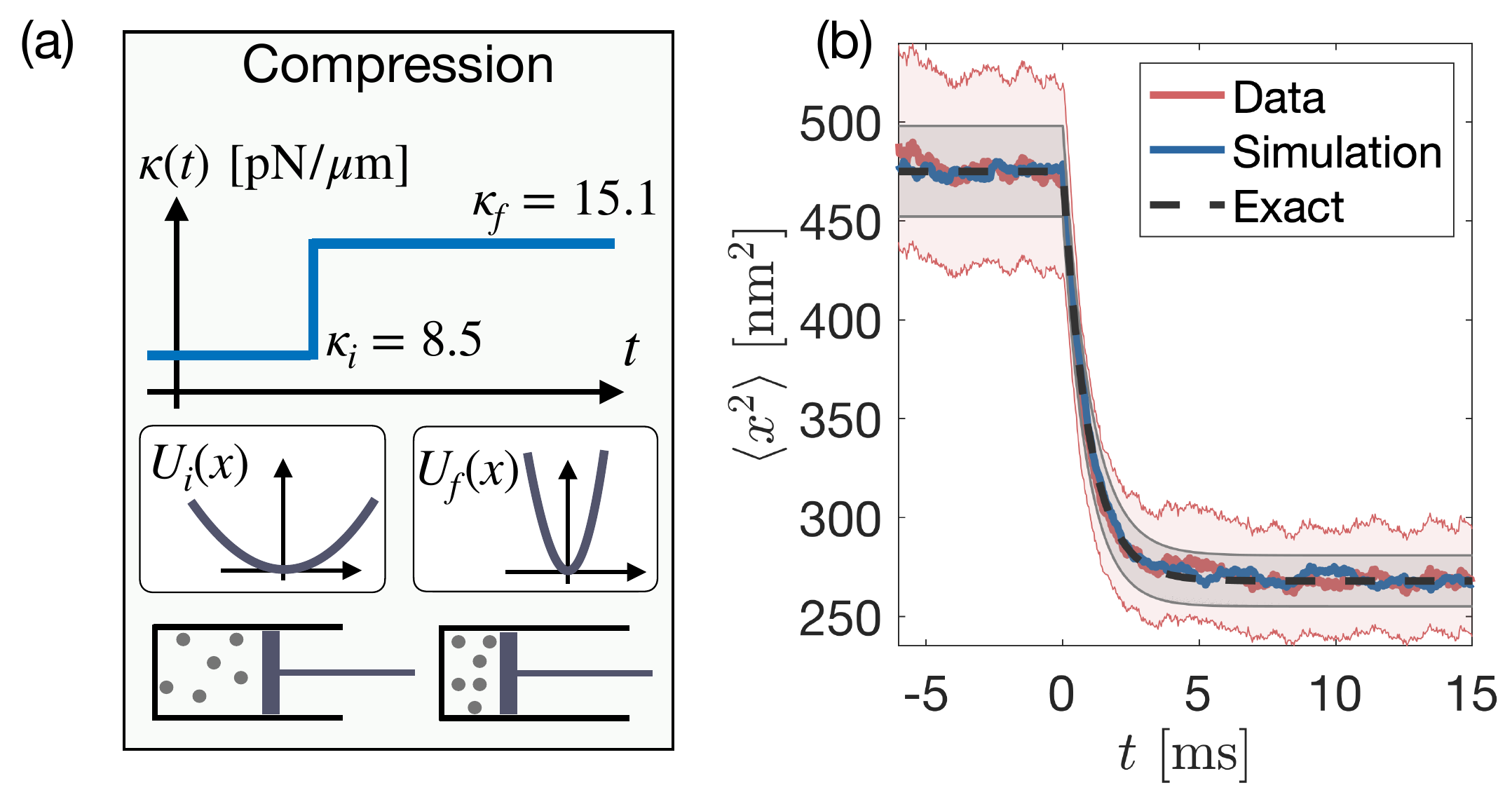}}
	\caption{\textbf{Stiffness change triggers NESS-to-NESS transformations.} (a) An illustration of a step-like change of stiffness $\kappa_\mathrm{i} \rightarrow \kappa_\mathrm{f}$, along with a schematic of the resulting potential, and a conceptual analogy to a compression experiment of an ideal gas.
    (b) Measured time-dependent ensemble-variance $\langle x^2(t) \rangle$ (red solid line), under a stationary noise $\eta(t)$ with correlation time $\omega_\mathrm{c}^{-1} = 1$ ms; numerical simulation results (blue solid line) and analytical evolution of the variance (black dashed line; see complete expression in the appendix) using the measured values of $\kappa$.
    }
	\label{fig:Step}
\end{figure}

To explore whether this emergent temperature serves as a thermodynamic variable, we perform abrupt stiffness changes $\kappa_{\rm i} \rightarrow \kappa_{\rm f}$, analogous to volume changes in macroscopic thermodynamics, as shown in Fig.~\ref{fig:Step}(a).
The protocol is applied sequentially to the microsphere, providing an ensemble of more than $14,000$ trajectories of duration $30~\mathrm{ms}$ each following the same protocol $\kappa(t)$.
The laser beam controlling the trap stiffness is regulated independently of the beam generating the nonthermal stochastic force. From a thermodynamic perspective, modulations of $\kappa(t)$ therefore perform work on the particle, while the stochastic forcing contributes—together with thermal noise—to heat exchange. Consequently, a step-like change of stiffness at fixed noise corresponds to a vertical shift between nonequilibrium steady states in the effective-temperature representation shown in Fig. \ref{fig:Schema}(i).

For all drivings studied, the position distribution $P(x,t)$ remains indistinguishable from a Gaussian profile with a zero mean at all times.
As such, instantaneous properties are fully characterized by the second moment $\langle x^2 (t) \rangle$.
We show in Fig.~\ref{fig:Step}(b) the time-dependent variance $\langle x^2(t) \rangle$ for the step-like change of stiffness shown above, under the influence of a colored noise $\eta(t)$ with correlation time $\omega_\mathrm{c}^{-1} = 2.5~\mathrm{ms}$.
For time $t>0$, it relaxes to a new value, corresponding to a new Boltzmann-like NESS distribution.
As stressed above, $\T = \T(\kappa)$ and during the transformation  $\kappa_{\rm i} \rightarrow \kappa_{\rm f}$, the variance does not change according to the equipartition with the same temperature.

State-dependent effective temperatures have previously been identified in active Ornstein–Uhlenbeck processes and related models \cite{SzamelPRE14, WoillezPRL20, Holubec2020, MartinPRE21, hecht2024define, Wiese2024}.
This explicitly contradicts equilibrium thermodynamics, which assumes that temperature is a property of the bath, independent of the system's state.
This should be distiguished from frequency-dependent effective temperatures \cite{zamponi2005fluctuation, zamponi2005generalized, dieterich2015single} defined through FRR, that where used in glassy systems \cite{crisanti2003violation, joubaud2009aging} and systems with mixed time-scales \cite{gnoli2014nonequilibrium}.
In this work, the temperature of Eq.~(\ref{Eq:Teff}) is not frequency dependent, and does not reconcile the FRR as we will show below (and in Appendix). Instead, motivated by a former approach~\cite{sorkin2024second}, we experimentally and analytically confirm the thermodynamic relevance of this state-dependent temperature for NESS-to-NESS transformations in active systems.

We now show that the stiffness-dependent effective temperature enables a consistent stochastic thermodynamic description of the NESS-to-NESS transformation.
Within the framework of stochastic thermodynamics \cite{Sekimoto1998,SekimotoBook,seifert2012,book:stochastic}, we express entropy production in terms of the nonequilibrium free energy.
Following Sekimoto’s approach, we identify a generalized work and heat from the Langevin equations and relate the entropy production to changes in work, free energy, and temperature. We then verify that the stiffness-dependent temperature satisfies the second law. 

The microsphere in our experimental system is continuously subjected to colored noise (Eq.~(\ref{Eq:Langevin}), with constant $\omega_\mathrm{c}$ and $\sigma_\eta$), reaching a NESS with $\kappa(t<0)=\kappa_\mathrm{i}$. It then undergoes an abrupt change of stiffness $\kappa(t>0)=\kappa_\mathrm{f}$, after which we allow the particle to reach the new NESS. Motivated by the classical second law of thermodynamics, describing the irreversibility of equilibrium-to-equilibrium transformation, we seek to quantify the generated entropy and thus work along this NESS-to-NESS transformation.

The Gaussian steady-state statistics allow for an equilibrium-like thermodynamic construction with a well-defined reversible limit, in which the energetic cost vanishes for infinitely slow protocols. The effective temperature $\T[\kappa(t)]$ depends on the control parameter. As shown above, applying a change in stiffness induces an immediate change in temperature, in contrast to isothermal equilibrium transformations \cite{book:stochastic,jarzynski1997nonequilibrium,crooks1999entropy}. This feature necessitates a modification of the standard stochastic thermodynamic formulation of the second law.

We define the stochastic system entropy and the nonequilibrium free energy as
\begin{align}
    \sigma_{\rm sys}(x(t),\kappa(t)) &= -k_\mathrm{B} \ln[P(x(t),\kappa(t))], \label{eq:SysEntropy}\\ 
    \mathcal{F}(\kappa(t)) &= -k_\mathrm{B} \T[\kappa(t)]\ln[\mathcal{Z}(\kappa(t))]\label{eq:free-energy}
\end{align}
where $\sigma_{\rm sys}(x(t),\kappa(t))$ is evaluated using the distribution $P(x(t),\kappa(t))$, \textit{i.e.} the steady-state distribution Eq.~(\ref{eq:effboltzmann}) corresponding to the instantaneous values of $\kappa(t)$ and $\T(t)$~\cite{hatano2001steady, Pires2023}. It does not correspond to the distribution of the time-dependent ensemble of $x(t)$ considered in the standard expression of $\sigma_{\rm sys}$ for isothermal processes~\cite{Seifert2005}.
For a given $\kappa$, $\mathcal{F}(\kappa(t))$ is the NESS free energy, and $\mathcal{Z}(\kappa(t)) = [2\pi k_\mathrm{B}\T[\kappa(t)]/\kappa(t)]^{1/2}$ is the partition function of the NESS distribution.

Taking the total differential of $\sigma_{\rm sys}$ along a trajectory yields
\begin{multline}
    d \sigma_{\rm sys}(x(t),\kappa(t))=\frac{x^2(t)}{2 \T[\kappa(t)]} d\kappa(t) - \frac{\kappa(t) x^2(t)}{2 \T^2(\kappa(t))} d\T(t)\\ +\frac{\kappa(t) x(t)}{\T[\kappa(t)]}\circ dx(t)- d\left(\frac{\mathcal{F}(\kappa(t))}{\T[\kappa(t)]}\right);
\end{multline}
as detailed in Appendix. This expression immediately gives rise to equilibrium-like expressions of heat (third term, $dq=-\kappa x\circ dx$, with the sign convention that a positive heat is dissipated in the bath) and change in free energy over temperature (fourth term) \cite{esposito2011second}. At the same time, the work (first term, $dw=(x^2/2)d\kappa$) over temperature is not the only contribution to entropy change, as changing the stiffness modifies in turn the effective temperature $\T[\kappa(t)]$. Therefore, we are motivated to combine (and divide by $k_{\rm B}$) the first two terms and define the generalized and normalized work-like quantity
\begin{equation}
    d\bar w= \frac{x^2}{2k_\mathrm{B}\T}d\kappa- \frac{\kappa x^2}{2k_\mathrm{B}\T^2}d\T,
    \label{Eq:WorkDef}
\end{equation}
which includes both the mechanical contribution $\sim d\kappa$ and the mechanically-induced thermal contribution $\sim d\T$ \cite{Rademacher2022, Pires2023}.

The total entropy production, $d\sigma_\mathrm{tot}=d\sigma_\mathrm{sys}+dq/\T$, now reads
\begin{equation}
    d\sigma_{\rm tot} = k_\mathrm{B} d \bar w - d \left(\frac{\mathcal{F}}{\T}\right)
\label{Eq:totalEntropy}
\end{equation}
with the sign convention that a positive work is a work applied on the system.
We verify experimentally and analytically that the average entropy production satisfies $\langle\sigma_\mathrm{tot}\rangle\geq0$ for all protocols, and that this bound is saturated in the quasistatic limit, $\langle\sigma_\mathrm{tot}\rangle=\mathcal{O}(\Delta\kappa^2)$ (see Appendix). In the following, we demonstrate that this formulation  leads to a modified Crooks FT for the accumulated generalized work $\bar w$.

\section*{Generalized work and a Crooks-like fluctuation relation}

As shown above, reconstructing a second-law–like inequality for NESS-to-NESS transformations requires a generalized definition of work that explicitly accounts for the state dependence of the effective temperature. This generalized work differs from the isothermal work defined in standard stochastic energetics \cite{Sekimoto1998}. Using the conventional definition in the presence of an active bath does not allow a generalization of the Jarzynski equality or the associated second law, as demonstrated in Ref.~\cite{szamel2023single}. Instead, the implicit evolution of the effective temperature must be taken into account \cite{Holubec2020, WoillezPRL20, MartinPRE21, hecht2024define, Wiese2024}.

This requirement is naturally satisfied within the equilibrium-like formulation developed here. The total work exchanged during a state-to-state transformation is obtained by integrating the generalized work increment $d\bar w$ over time (see Appendix).
For the step-like change in stiffness from $\kappa(0) = \kappa_\mathrm{i}$ to $\kappa(t>0) = \kappa_\mathrm{f}$ it takes a simple form, 
\begin{equation}
    \bar w(t) = \frac{1}{2}\left(\frac{\kappa_\mathrm{f}}{k_\mathrm{B} \T(\kappa_\mathrm{f})} - \frac{\kappa_\mathrm{i}}{k_\mathrm{B} \T(\kappa_\mathrm{i})}\right) x^2(0),
    \label{Eq:stepWork}
\end{equation}
which depends quadratically on the stochastic position $x^2(0)$ at the instant of the transformation.
Using Eq.~(\ref{Eq:stepWork}), we derive the probability distribution of generalized work for the forward protocol, $p(\bar w)$ (expression given in Appendix). To quantify irreversibility, we also consider the backward protocol, in which the stiffness changes $\kappa_\mathrm{f} \rightarrow \kappa_\mathrm{i}$. The corresponding work distribution is denoted $\tilde p(\bar w)$. These two distributions of generalized work satisfy the relation
\begin{equation}
    \ln\left[ \frac{p(\bar w)}{\tilde p(-\bar w)} \right] = \bar w - \left( \frac{\mathcal{F}(\kappa_{\rm f})}{k_\mathrm{B} \T(\kappa_{\rm f})} -  \frac{\mathcal{F}(\kappa_{\rm i})}{k_\mathrm{B} \T(\kappa_{\rm i})}\right).
    \label{Eq:CFT}
\end{equation}
This equality is a Crooks-like FT \cite{crooks1999entropy, esposito2010three} which is now valid for transformation starting and ending at NESSs with arbitrarily strong active driving.

\begin{figure}
    \centering
    \includegraphics[width=1\linewidth]{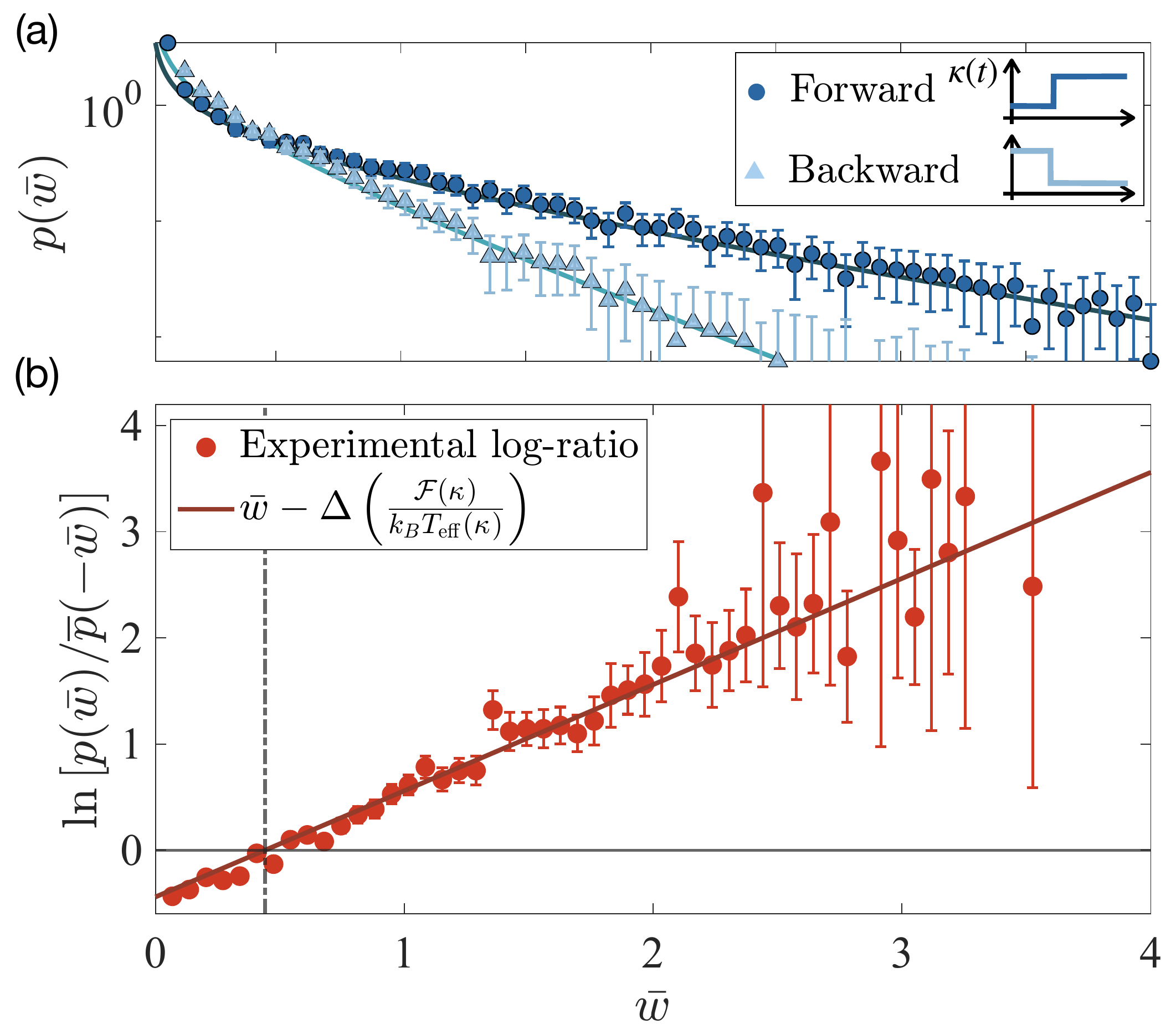}
    \caption{\textbf{Fluctuation theorem for the generalized work.} (a) Generalized work distributions for both forward transformation (compression, dark blue circles) and backward transformation (expansion, light blue triangles) for a system driven by colored noise. Solid lines corresponds to the exact results (see Appendix) using the stiffness $\kappa_{\rm i}$ and $\kappa_{\rm f}$ measured in the absence of noise (at thermal equilibrium) and temperatures $\T(\kappa_\mathrm{i})$ and $\T(\kappa_\mathrm{f})$.
    (b) Fluctuation theorem for the generalized work confirmed using the experimental data (red circles) on top of the exact result Eq.~(\ref{Eq:CFT}) (red solid line). The expected normalized free-energy difference $\Delta (\mathcal{F}/k_{\rm B}\T)$ is underlined (dot-dashed vertical line). It agrees with the value of work for which the log-ratio crosses $0$, demonstrating that finite-time measurement of $\bar w$ allows to probe the free-energy differences (see Fig.~\ref{fig:DeltaF} in Appendix).}
    \label{fig:CFT}
\end{figure}

To confirm this generalized Crooks FT, we compute the generalized work, Eq.~(\ref{Eq:stepWork}), along the ensemble of experimental trajectories obtained for abrupt compression $\kappa_\mathrm{i} \rightarrow\kappa_\mathrm{f}$.
The reverse expansion $\kappa_\mathrm{f}\to\kappa_\mathrm{i}$ is then applied, yielding an equal number of backward trajectories.
The time duration after each change of stiffness is much longer than $\omega_\mathrm{c}^{-1}$, $\gamma/\kappa_\mathrm{i}$, and $\gamma/\kappa_\mathrm{f}$, such that NESS is indeed attained at the end of each repetition.

Figure~\ref{fig:CFT}(a) shows the measured distribution of generalized work in both the forward and backward transformations under colored noise with constant correlation time $\omega_c^{-1} = 2.5~\rm{ms}$.
Both distributions exhibit a combination of a power-law decay with an exponential tail, in very good agreement with the analytical result.
The log-ratio of both work distributions validates experimentally the generalized Crooks-like FT (Fig.~\ref{fig:CFT}(b)) (all cases of $\omega_{\rm c}$ are shown in Fig.~\ref{fig:AllFT} in the Appendix).
This simple relation highlights the effectiveness of applying equilibrium-like approaches to describe this far-from-equilibrium system \cite{boudet2025non}.

Our framework deliberately excludes the \textit{housekeeping} cost required to maintain the NESS under continuous driving \cite{oono1998steady}. Including this contribution would lead to diverging entropy production over long times and preclude a reversible-like limit, as discussed in Refs.~\cite{fodor2021active, davis2024}. Instead, we focus on the thermodynamic cost of externally controlled parameter changes, treating the microscopic driving mechanism as cost-free. This separation is natural in many active systems, where energy input (e.g., chemical fuel) is distinct from mechanical actuation.
Within this perspective, incorporating the effect of the nonequilibrium bath through the effective temperature $\T$ enables us to establish strong constraints on the energetics of NESS-to-NESS transformations.

The presence of a fluctuation relation for the work exchanged, and consequently for total entropy production, has direct practical implications: it provides a way to measure the free-energy difference (\textit{i.e.} the minimal work reached for infinitely slow transformations) from a finite-time transformation. This demonstrates the predictive value of the state-dependent effective temperature $\T[\kappa]$ as a thermodynamic variable governing energy exchange in active and driven systems, with Gaussian fluctuations.

Finally, while the effective temperature captures the symmetry of energy exchanges during NESS-to-NESS transformations, it does not enforce equilibrium-like dynamical relations. We therefore turn next to fluctuation–response measurements to examine how nonequilibrium memory effects show up in the system’s temporal dynamics.

\section*{A fluctuation-response relation for variance}

At thermal equilibrium, FRR relates the spontaneous stationary fluctuations of a system, to its response to a perturbation \cite{kubo1966fluctuation}.
Deviations from the FRR can reveal excess fluctuations due to an active driving force \cite{dinis2012fluctuation, dieterich2015single, turlier2016equilibrium, marconi2017heat, Sandford2017, Goerlich2022} or hidden sources of dissipation such as friction at macroscopic scale \cite{boriskovsky2024fluctuation}.
Thus, in both instances, it can serve as a diagnostic tool for identifying the system's nonequilibrium nature as well as the consequence of non-Markovianity induced by correlated noise \cite{maggi2014generalized, Sandford2017, Goerlich2022, engbring2023nonlinear} or other hidden variables \cite{WilSok17}.

To probe the dynamics during transformations between distinct NESS, we reinterpret the same stiffness-change experiment used above to validate the Crooks fluctuation theorem. We now analyze how the system’s response to the stiffness perturbation compares to spontaneous fluctuations measured in the final steady state. This is an FRR approach, which differs from standard constant-force perturbations \cite{dieterich2015single, boriskovsky2024fluctuation}, as they do not probe transitions between thermodynamically distinct states (with changing free energy).

Along this stiffness-perturbation, we  monitor the response $\mathcal{R}(t) = \langle x^2(t)\rangle - \langle x^2 \rangle_{\rm f}$. This response is then compared to the unperturbed fluctuations of $x^2$ in the final steady state under $\kappa_\mathrm{f}$, quantified by its positional autocorrelation function.
Using Wick's theorem in this Gaussian case, the correlations of $x^2$ can be further related to the square of the positional correlation function $C_{xx}(t) \equiv \langle x(t) x(0) \rangle_{\rm f}$, leading to the following FRR
\begin{equation}
    \mathcal{R}(t) = \frac{\Delta\kappa}{k_\mathrm{B} T}\left(1 + \frac{\Delta\kappa}{\kappa_\mathrm{i}}\right) C_{xx}^2(t),
    \label{Eq:FRR}
\end{equation}
where $\Delta \kappa = \kappa_\mathrm{f} - \kappa_\mathrm{i}$ is the amplitude of the perturbation.
This result constitutes an FRR for the variance, valid for a perturbation of arbitrary amplitude, in a thermal bath at temperature $T$ and reduces to linear-response theory for $\Delta\kappa/\kappa_\mathrm{i}\ll1$ \cite{kubo1966fluctuation}. For another nonlinear FRR that also holds for athermal Markovian systems, see Ref.~\cite{engbring2023nonlinear}.

\begin{figure}[t!]
    \centering
    \includegraphics[width=0.92\linewidth]{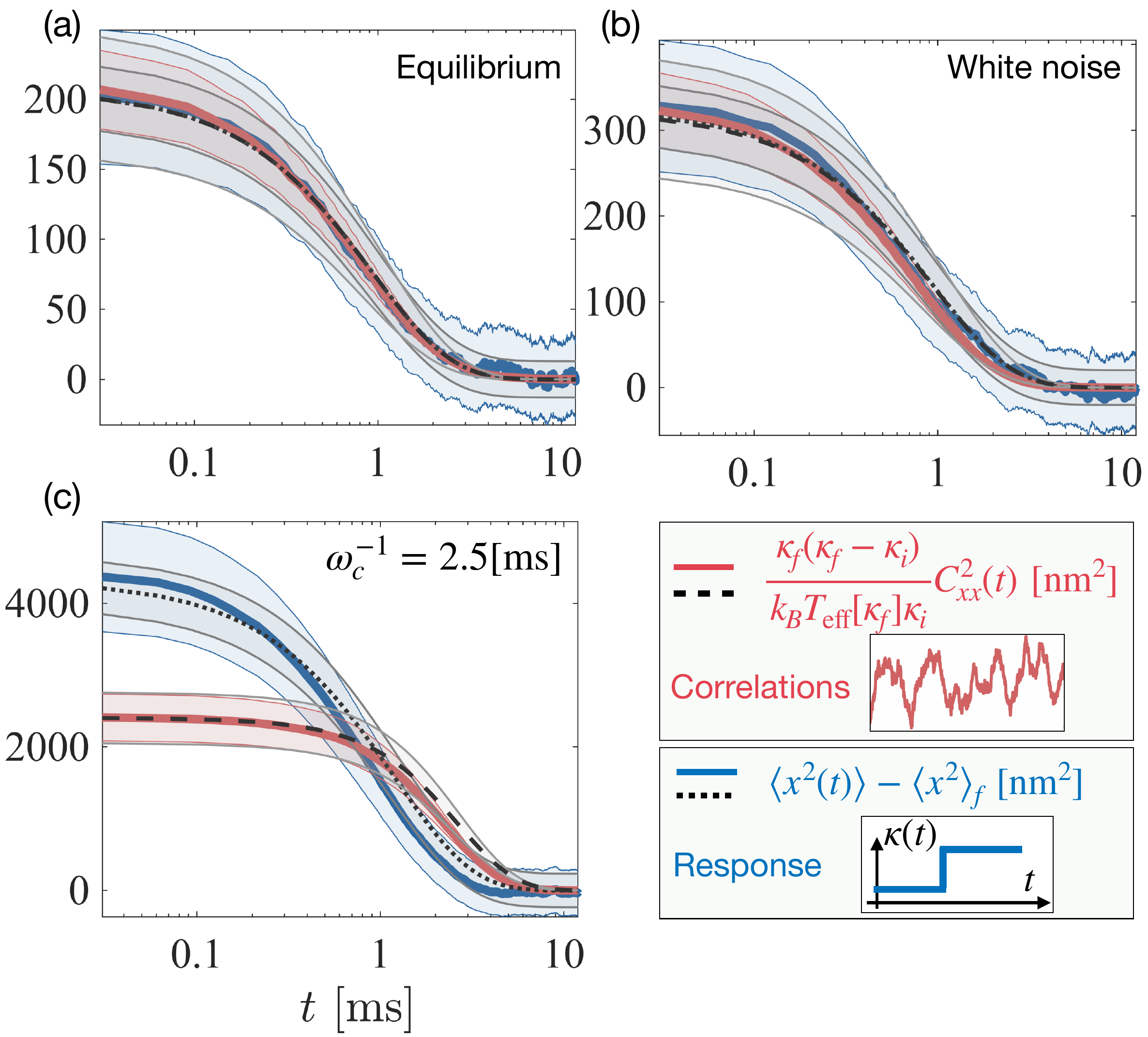}
    \caption{\textbf{Fluctuation-response relation under stiffness change.}
    (a) Verification of the variance-FRR at thermal equilibrium, showing excellent agreement between the response function (experimental data, blue solid line; analytical prediction, black dotted line) and the squared correlation function multiplied by the prefactor in Eq.~(\ref{Eq:FRR}) (experiment, red solid line; analytical prediction, black dashed line).
    The time-dependent second moment and the correlation function are given by Eqs.(\ref{eq:NeqVar}) and (\ref{eq:NeqCxx}) in the Appendix, respectively. Shaded regions around the experimental curves indicate a $\chi^2$ test with a $3\sigma$ confidence interval and include the propagation of calibration uncertainties (see Appendix). Shaded regions around the analytical curves reflect uncertainty propagation from experimentally measured parameters.
    (b) Verification of a generalized variance-FRR under white-noise driving, using the effective temperature $\T$ in the prefactor of the correlation function.
    (c) Violation of the variance-FRR under colored-noise driving with $\omega_{\mathrm{c}}^{-1}=2.5~\mathrm{ms}$, using $\T(\kappa_{\mathrm{f}})$ in the prefactor.
    }
    \label{fig:FRR}
\end{figure}

We first verify Eq.~(\ref{Eq:FRR}) experimentally at thermal equilibrium. Fig.~\ref{fig:FRR}(a) shows excellent agreement between the measured response $\mathcal R(t)$ and the appropriately rescaled squared correlation function $C^2_{xx}(t)$, confirming the validity of the variance-FRR for equilibrium-to-equilibrium transformations. Deviations from this relation therefore provide a direct diagnostic of nonequilibrium dynamics. This motivates the following question: can Eq.~(\ref{Eq:FRR}) be restored by replacing $T$ with the effective temperature $\T(\kappa_{\rm f})$ characterizing the final NESS?

As shown below, answering this question with our experimental data lifts a degeneracy between two distinct nonequilibrium drivings.
For systems driven by additional white noise, the variance-FRR is recovered by substituting $T$ with $\T$, reflecting the Markovian nature of the dynamics. In contrast, for systems driven by colored noise, the FRR cannot be restored due to the distinct temporal evolutions of the response and correlation function.
This highlights the role of non-Markovianity in creating a nonequilibrium state that dynamically differs from effective equilibrium.

Fig.~\ref{fig:FRR}(b) shows that under white-noise driving, rescaling the correlation function by the effective temperature $\T$ restores the variance-FRR. Although the system is driven out of equilibrium with respect to the thermal bath, response and correlations share the same time dependence. The dynamics are therefore indistinguishable from those of an equilibrium system at a higher temperature \cite{martinez2013effective, Rademacher2022, Pires2023}.

By contrast, Fig.~\ref{fig:FRR}(c) demonstrates that for colored-noise driving with correlation time $\omega_c^{-1} = 2.5~\rm{ms}$, rescaling by 
$\T(\kappa_{\rm f})$ fails to restore the FRR (all cases of $\omega_{\rm c}$ are shown in Fig.~\ref{fig:AllFRR} in the Appendix). Response and correlations differ both in amplitude and time dependence, reflecting the presence of memory effects introduced by the colored noise. In this case, no constant, or even state-dependent, temperature can reconcile the two. Restoring FRR requires a frequency-dependent temperature \cite{cugliandolo2011effective}, which we discuss in Appendix. Such temperatures, however, lack a clear thermodynamic interpretation in systems with mixed time scales \cite{gnoli2014nonequilibrium, boudet2025non}.

Thus, while the effective temperature $\T(\kappa)$ successfully characterizes the thermodynamics of NESS-to-NESS transformations and underpins the fluctuation relation Eq.~(\ref{Eq:CFT}), it does not enforce equilibrium-like dynamical relations. The violation of the variance-FRR Eq.~(\ref{Eq:FRR}) therefore provides a complementary dynamical signature of nonequilibrium and non-Markovian behavior.

\section*{Conclusion}
Thermodynamics provides a unifying framework for transformations between equilibrium states, yet extending this structure to nonequilibrium systems remains challenging.
Here we show that a consistent thermodynamic description can emerge when focus is shifted from individual nonequilibrium steady states to the transformations between them.
Using an optically trapped colloidal particle driven by a controlled, active-like stochastic force, we demonstrate that a state-dependent effective temperature, defined by stationary fluctuations, governs NESS-to-NESS transformations. When combined with a generalized definition of work that accounts for its stiffness dependence, this temperature satisfies a second-law–like inequality and leads to a Crooks-like fluctuation theorem, preserving key symmetries of equilibrium thermodynamics far from equilibrium.

At the same time, our results delineate clear limits to this description. A fluctuation–response relation for the variance holds at equilibrium and under white-noise driving but fails under colored-noise forcing, reflecting the non-Markovian nature of the reduced $x(t)$ dynamics. Thus, the effective temperature that characterizes the thermodynamics of transformations does not restore equilibrium-like dynamical response. This separation highlights that the thermodynamic and dynamical notions of temperature do not necessarily coincide in nonequilibrium systems, and that thermodynamic consistency can persist even when response relations are violated. Although our study focuses on harmonic confinement and exponentially correlated Gaussian noise, both ingredients are broadly relevant: harmonic potentials describe local behavior near general energy minima, and this type of colored noise captures essential features of many active environments \cite{fily2012athermal, maggi2014generalized}.

Beyond this specific setting, the transition-based framework developed here provides a basis for identifying when nonequilibrium systems admit thermodynamic structure and when they fundamentally depart from equilibrium analogies. For weak active noise, approximate treatments such as the unified colored-noise approximation suggest that similarly simple thermodynamic descriptions may extend to non-harmonic potentials \cite{jung1987dynamical, hanggi1994colored, li1995bistable}, while retaining access to rich nonequilibrium dynamics. Setting aside a full account of housekeeping energetics, this framework offers two immediate directions of application: understanding and constraining energetic costs of mesoscopic biological processes, such as driven steps of molecular motors \cite{ariga2021noise}, and guiding the design and optimization of mesoscopic engines operating in nonequilibrium environments \cite{krishnamurthy2016micrometre, Holubec2020}.

\acknowledgements{
R.G. and YR acknowledge support from the European Research Council (ERC) under the European Union’s Horizon 2020 research and innovation program (Grant Agreement No. 101002392). R.G. acknowledges support from the Mark Ratner Institute for Single Molecule Chemistry at Tel Aviv University.
B.S. acknowledges support from the Princeton Center for Theoretical Science.
This work is also part of the Interdisciplinary Thematic Institute QMat of the University of Strasbourg, CNRS, and Inserm, supported by the following programs: IdEx Unistra (Grant No. ANR- 10-IDEX-0002), SFRI STRATUS Project (Project No. ANR-20-SFRI-0012), and USIAS (Grant No. ANR-10- IDEX- 0002-02), under the framework of the French Investments for the Future Program.
BL would like to thank Deutsche Forschungsgemeinschaft for their support (DFG grant LI 1046/10-1).
C.G. acknowledges the support of the French Agence Nationale de la Recherche (ANR), under grant  ANR-23-CE30-0042 (project FENNEC)
Y.R. acknowledges support from the Israel Science Foundation (grants No. 385/21).
}


%

\appendix

\section{Experimental setup, calibration and evaluation of the errors}
\label{App:Exp}

\begin{figure}[h!]
	\centerline{\includegraphics[width=0.9\linewidth]{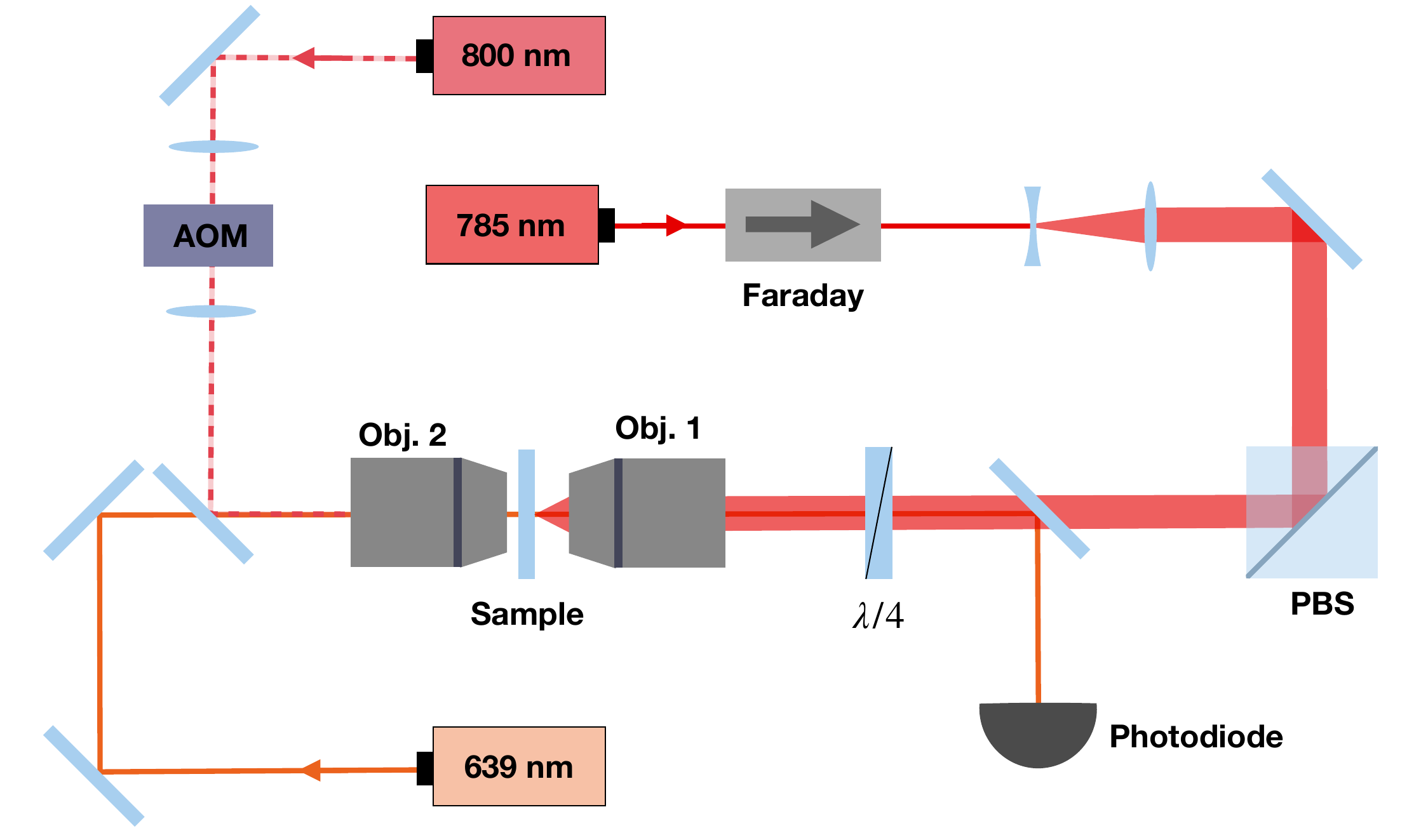}}
	\caption{Simplified view of the optical trapping setup. The sphere is suspended in water inside the \textit{Sample} cell inserted between the two objectives Obj1 and Obj2. The $785$ nm trapping beam is drawn in red. The $800$ nm beam used to apply  radiation pressure is shown in purple. The intensity of this beam is controlled by the acousto-optic modulator (AOM). The instantaneous position of the trapped bead is probed using the auxiliary $639$ nm laser beam, drawn in orange, whose scattered signal is sent to a high-frequency photodiode.
		}
	\label{fig:AppSchema}
\end{figure}

\textit{Experimental platform -}
Our setup (sketched in Fig.~\ref{fig:AppSchema}) consists in optically trapping, in a harmonic potential, a single dielectric bead ($3~\rm{\mu m}$ polystyrene sphere) in a fluidic cell filled with deionized water at room temperature $T = 296$ K.
The harmonic potential is induced by focusing inside the cell a linearly polarized Gaussian beam ($785$ nm, CW $110$ mW laser diode, Coherent OBIS) through a high numerical aperture objective (Nikon Plan Apo VC, $60\times$, NA$=1.20$ water immersion, Obj1 on Fig. \ref{fig:AppSchema}). An additional force in the form of radiation pressure is applied to the sphere using a time-dependent fraction of the light-beam emitted by an additional high-power laser ($800$ nm, CW $5$ W Ti:Sa laser, Spectra Physics 3900S).
The intensity of this radiation pressure beam is controlled by an acousto-optic modulator (Gooch and Housego 3200s, AOM on Fig. \ref{fig:AppSchema}) using a digital-to-analogue card (NI PXIe 6361) and a \textsc{python} code.
It is sent on the micrsophere as a thin beam, strongly underfilling the aperture of a lower-NA objective (Nikon Plan Fluor Extra Large Working Distance, $60\times$, NA$=0.7$, Obj2 in Fig.~\ref{fig:AppSchema}) in order to prevent additional gradient forces.

The instantaneous position $x(t)$ of the sphere along the optical axis is measured by recording the light scattered off the sphere of a low-power $639$ nm laser (CW $30$ mW laser diode, Thorlabs HL6323MG), sent on the bead via the second, low-NA objective. The scattered light is collected by Obj1 and recorded by a photodiode ($100$ MHz, Thorlabs Det10A).
The recorded signal (in V/s) is amplified using a low noise amplifier (SR560, Stanford Research) and then acquired by an analog-to-digital card (NI PCI-6251). The signal is filtered through a $0.3$ Hz high-pass filter at 6 dB/oct to remove the DC component and through a $100$ kHz low-pas filter at 6 dB/oct to prevent aliasing. The scattered intensity varies linearly with the position of the trapped bead $x(t)$ for small enough displacements and we make sure to work in the linear response regime of the photodiode so that the recorded signal is linear with the intensity, resulting in a voltage trace $v(t)$ well linear with the position $x(t)$ of the microsphere in the trap.\\

\textit{Calibration of $x$ and $\kappa$, errors estimation -} First we perform an equilibrium steady-state recording in a trap of constant stiffness to evaluate the \textit{volt to meter} coefficient $\beta$ that relies on the linearity detailed above.
The fit of the power-spectral density (PSD) of the recorded trajectory with a Lorentzian distribution leads to $\beta = 0.13 \pm 0.01~\rm{V/\mu m}$ with an error of $7\%$.
This error stems from a combination of the error in the PSD fit and the $5\%$ uncertainty on the microsphere's radius (via the viscous drag $\gamma$).
Next, we perform a transformation between two stiffnesses $\kappa_{\rm i}$ and $\kappa_{\rm f}$ in the absence of driving noise.
The system therefore reaches equilibrium in both state and equipartition allows to estimate both stiffnesses as $\kappa_{\rm i} = k_{\rm B}T / \langle x^2 \rangle_i $ and resp. $\kappa_{\rm f}$.
The combination of the error on the coefficient $\beta$ and the statistical error on the variances (measured via a $\chi2$-test) leads to an error of $15\%$ on the measured stiffnesses.
Comparing the measured variances to the integral of the fitted PSD allows to unveil a systematic overestimation of measured variance (arising from spurious high frequency electronic noise) of $\epsilon = \langle x^2 \rangle / \int S_{xx}[\omega] d\omega \approx 7\%$.
When comparing experimental variances to their analytical expression, we use this systematic correction.
The error in variance leads to the shaded area around the experimental data (red shaded area in Fig.~\ref{fig:Step}(b); red and blue shaded area in Fig.~\ref{fig:FRR}) while the error in $\kappa$ is propagated in the associated analytical results (gray shaded area in Fig.~\ref{fig:Step}(b) and Fig.~\ref{fig:FRR}).
Both stiffness and measured variance determines the effective temperature $\T$ and the propagation of the errors lead to an significative uncertainty of $13.5\%$ on the effective temperatures (see errorbars in Fig.~\ref{fig:Step}(c)).
Errors $\delta \bar w$ in the work, derived from the ensemble of trajectories, are determined by errors of $\kappa$, $\T$ and $\beta$ as visible in Eq.~(\ref{Eq:stepWork}).
They are propagated to the distributions $p(\bar w)$ by assuming a Gaussian distribution of errors $\mathcal{N}(\bar w, \delta \bar w^2)$.
For a discrete histogram, a measured work $\bar w_i$ has a probability $p_{ij} = \int_{a_j}^{b_j} \mathcal{N}(\bar w_i, \delta \bar w_i^2)$ to fall within the bin $j$ with boundaries $a_j$ and $b_j$.
The error on the distribution is then given by $\delta p(\bar w) = \sum_{i,j} p_{ij}(1-p_{ij})$ (see Fig.~\ref{fig:CFT}(a)).
This error is finally propagated to the log-ratio of work histograms (see Fig.~\ref{fig:CFT}(b)).

\begin{figure}[h!]
	\centerline{\includegraphics[width=1\linewidth]{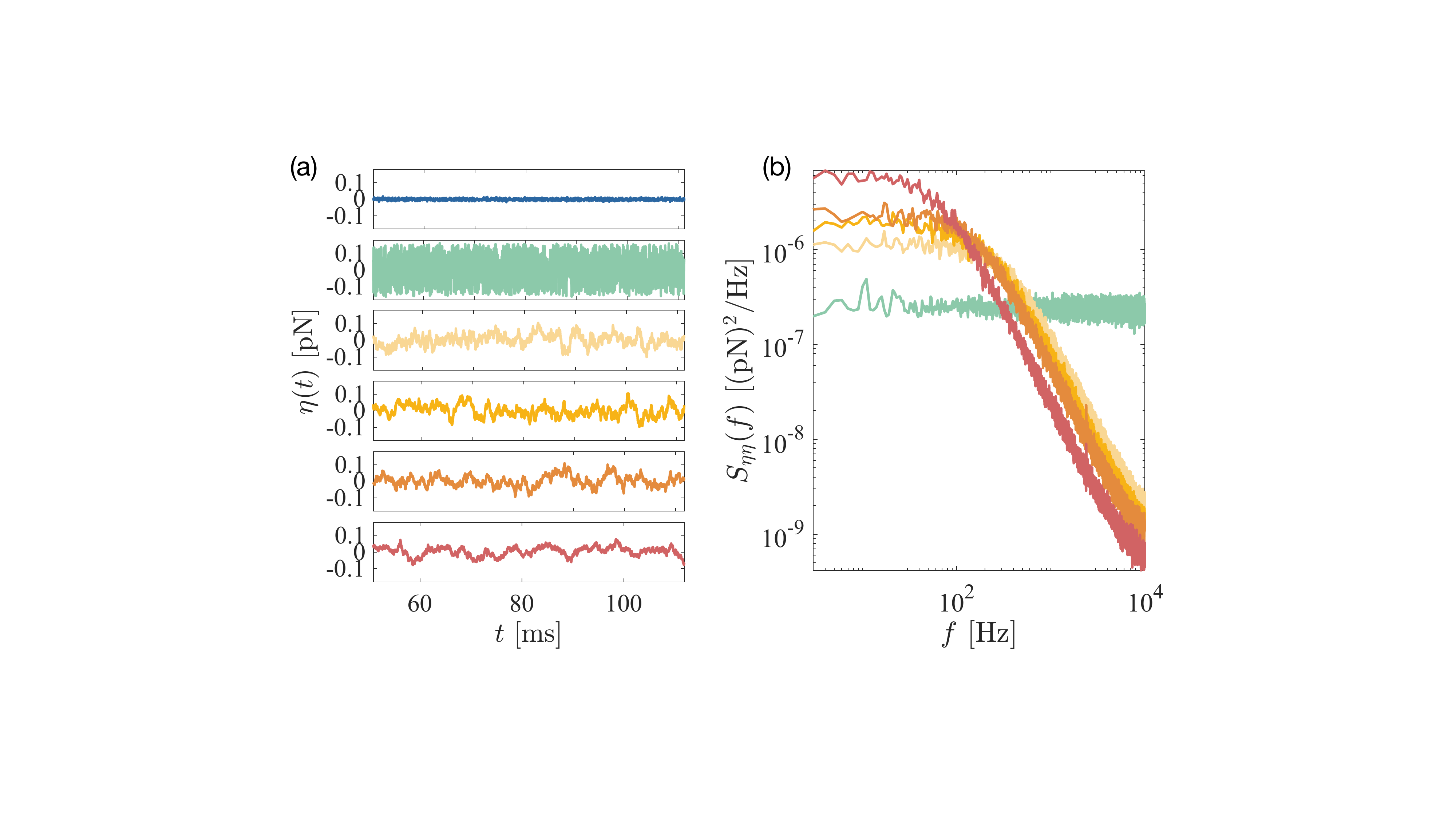}}	
	\caption{
    (a) snapshot of the noises $\eta(t)$ sent to the microsphere, in $\rm{p N}$. We show the measured DC signal (blue line), white noise (light green line) and all colored noises (yellow to red for $\omega_c^{-1} = 0.5$ to $2.5$ ms).
    (b) Power spectral density of each recorded noise. In the case of the DC signal (blue line), it only displays the electronic acquisition noise and spurious signals, the white noise generated up to $2^{15}~\rm{Hz}$ (light green line) is flat along the measured bandwidth and the various colored noises display a Lorentzian profile.
		}
	\label{fig:NoisePSD}
\end{figure}

\textit{Calibration of the external force amplitude ---} In the Langevin equation (Eq.~(\ref{Eq:Langevin})), the external noise $\eta(t)$ takes the form of an external force.
The absolute amplitude of this force depends of the coupling between the time-dependent light-intensity and the position degree of freedom of the trapped microsphere.
It is possible to calibrate this coupling, and therefore the intensity of the applied external force, from the effect of $\eta(t)$ on the recorded trajectories $x(t)$.
To do so, we rely on the stationary variance $\langle x^2 \rangle$, which can be derived via the integration of the power spectral density of position $S_{x}[\omega]$ which reads
\begin{equation}
    S_{x}[\omega] = \frac{1}{\omega_0^2 + \omega^2} \left( 2D + \frac{S_{\eta}[\omega]}{\gamma^2} \right)
\end{equation}
where $S_{\eta}[\omega]$ is the power spectral density of the driving force $\eta(t)$.
In the case of a white noise, the force obeys $\langle \eta(t) \eta(t) \rangle = \sigma_\eta^2 \delta(t-s)$, (with amplitude $\sigma_\eta$ having the dimension of a force multiplied by the square root of time).
Importantly here, the pseudo white-noise is generated only up to a finite cutoff pulsation $\omega_{\rm gen} = 2\pi \times 2^{15}~\rm{s^{-1}}$, leading to $S_{\eta}[|\omega| < \omega_{\rm gen}] = \sigma_\eta^2$ and $S_{\eta}[|\omega| > \omega_{\rm gen}] = 0$.
The positional variance then reads
\begin{equation*}
    \begin{aligned}    
        \langle x^2 \rangle &= \frac{D}{\omega_0} + \frac{1}{\gamma^2 \pi} \int_{-\omega_{\rm gen}}^{\omega_{\rm gen}} \frac{\sigma_\eta^2}{\omega_0^2 + \omega^2} d\omega\\
        & = \frac{D}{\omega_0} + \frac{\sigma_\eta^2}{\gamma^2 \omega_0\pi} \rm{arctan}\left[\frac{\omega_{\rm gen}}{\omega_0} \right]
    \end{aligned}
\end{equation*}
where the first term $D/\omega_0 = k_{\rm B}T/\kappa$ corresponds to the equilibrium equipartition second-moment and comes from the integration of the Lorentzian over all space.
The second term accounts for the effect of the external force with finite bandwidth.
Inverting this relation allows to derive $\sigma_\eta^2$ from the measured value of $\langle x^2 \rangle$.
In Fig.~\ref{fig:NoisePSD}(a) (green line in the second panel), we show a snapshot of the noise $\eta(t)$ sent as a force onto the microsphere, calibrated as $\eta(t) = \sqrt{\sigma_\eta^2 \omega_{\rm acq}} \bar \eta(t)$, where $\omega_{\rm acq}$ is the inverse of the acquisition time-increments and $\bar \eta(t)$ is a the normalized recorded noisy light-intensity.
In Fig.~\ref{fig:NoisePSD}(b, green line) we show the respective power spectral density which flat spectrum over all the frequency regime probed by our experiment.

In the case of an exponentially correlated noise, we have $\langle \eta(t) \eta(s) \rangle = \sigma_\eta^2 e^{-\omega_c |t-s|}$ (and $\sigma_\eta$ has the dimension of a force).
In that case, $S_\eta[\omega] = 2 \sigma_\eta^2 \omega_c /(\omega_c^2 + \omega^2)$ and the variance reads
\begin{equation}
    \langle x^2 \rangle = \frac{D}{\omega_0} + \frac{ \sigma_{\eta}^2}{\kappa\gamma(\omega_c + \omega_0)}.
\end{equation}
This relation can be inverted as well to obtain $\sigma_\eta^2$ from the measured $\langle x^2 \rangle$. 
In Fig.~\ref{fig:NoisePSD}(a) (yellow to red lines in the lower panels) we show the calibrated noise $\eta(t) = \sigma_\eta \bar \eta(t)$ and in Fig.~\ref{fig:NoisePSD}(b) we show the respective spectral densities.
The PSD of the colored noise present the typical Lorentzian profile of an exponentially correlated process.
Our choice of normalization of the colored noise ensures that all colored noises have the same finite-bandwidth variance : the integral of their PSD on the recorded bandwidth is constant.

Inserting our definition of the effective temperature $\T = \kappa\langle x^2 \rangle/k_{\rm B}$ in the above equation leads to the relation $\T(\kappa) = T \left( 1 + \frac{\sigma_\eta^2}{\gamma D(\gamma \omega_\mathrm{c} + \kappa)} \right)$ given in Eq.~(\ref{Eq:Teff}) in the main text, revealing the state-dependent nature of $\T$.

\begin{figure}[h!]
	\centerline{\includegraphics[width=1\linewidth]{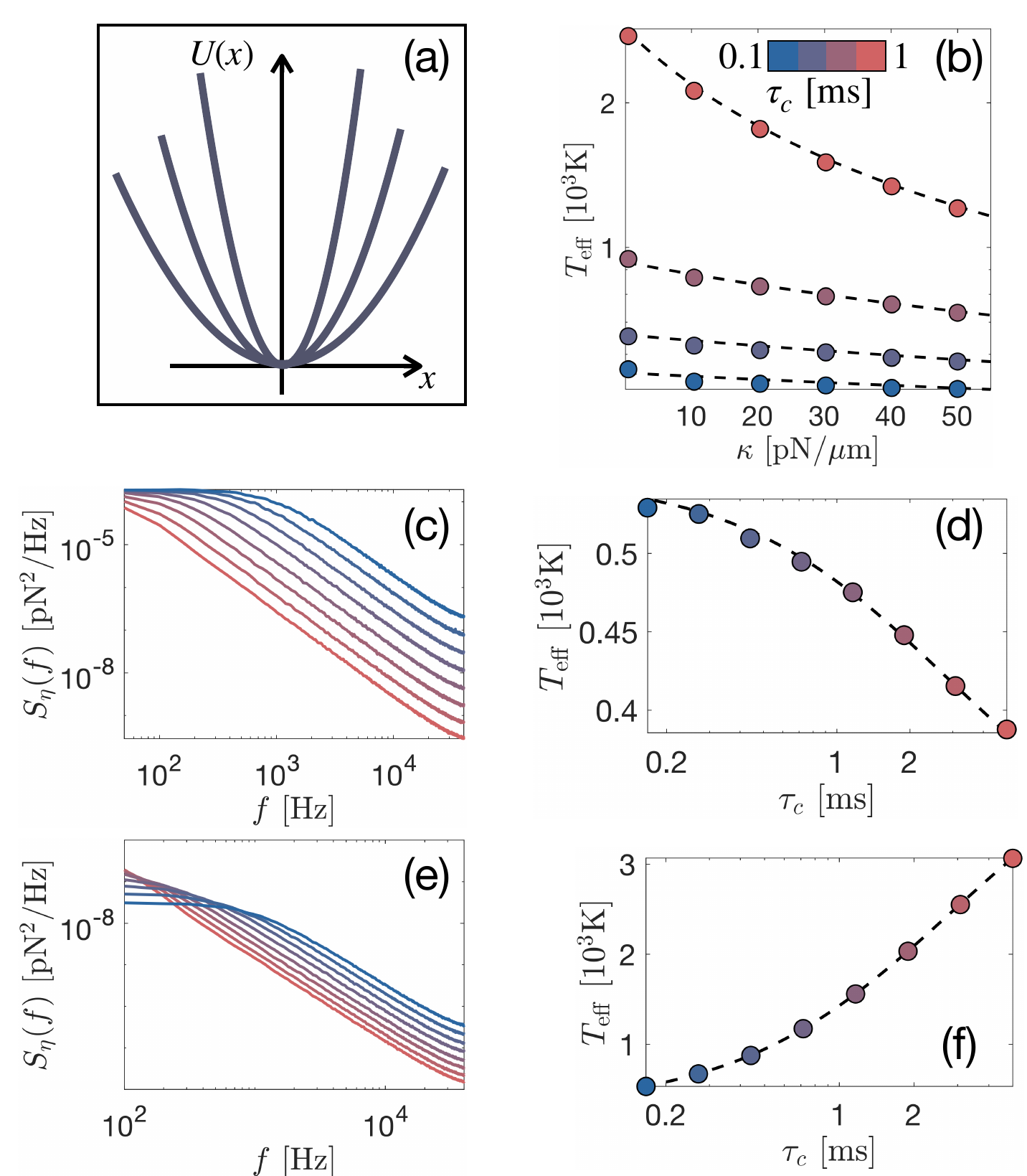}}
	\caption{(a,b) Effective temperature $\T$ measured with numerical simulations as a function of the stiffness $\kappa$ of the optical trap for various noise correlation time $\tau_c$. The solid line corresponds to the analytical result Eq.~(\ref{Eq:Teff}).
    (c) PSD of the driving noise $\eta(t)$ using a constant amplitude normalization.
    (d) associated effective temperature $\T$
    (e) PSD of the driving noise $\eta(t)$ for the same correlation times as (c), but using the constant variance normalization, as is done in the main text (fixing the variance $\sigma_\eta^2$ to match the variance of the smallest $\tau_c$ is panel (c))
    (f) associated effective temperature.
		}
	\label{fig:AppNormalization}
\end{figure}

In Fig.~\ref{fig:AppNormalization}, we show with numerical simulations the dependence of the effective temperature to the stiffness of the optical trap, complementing the data shown in Fig.~\ref{fig:Step}(c) in the main text.
This confirms that $\T$ decreases as the stiffness increases, due to the decreased mechanical coupling to the bath.
An important aspect when comparing colored noise of different correlation time $\tau_c$, is the choice of normalization.
Two options when varying $\tau_c$ are (\textit{i}) keeping the variance of the noise $\sigma_{\eta}^2 = \langle \eta^2 \rangle = \int S_{\eta}(f) df$ constant, or (\textit{ii}) keeping the noise amplitude $\sigma_{\eta}^2 \tau_c = S_\eta(0)$ constant.
We have followed the first choice in this work, motivated by the following argument: keeping the variance constant maintains the global power injected in the system by the driving noise, only modifying its spectral repartition. We believe that it is consistent with change in correlation time in colored noise in natural systems, with constant energy supply.
The second choice has been studied for instance in Ref.~\cite{saha2023information}.
In Fig.~\ref{fig:AppNormalization}(c,d) and (e,f) we comparatively study the consequences of both choices of the evolution of the effective temperature defined via the generalized equipartition.
It shows that the constant amplitude normalization leads to a decreasing temperature as $\tau_c$ increase.
An infinite temperature would be reached in the white noise limit.
In contrast, the constant variance normalization leads to an increasing temperature.
We also note that, for the same span of $\tau_c$, the constant variance normalization leads to a significantly larger range of temperature.

\section{Proof of second law of thermodynamics with equipartition temperature}
\label{App:SecondLaw}

In this section, we show that the effective temperature of Eq.~(\ref{Eq:Teff}) gives rise to the second law of thermodynamics for AOUPs. First, we recall that the temperature is stiffness dependent. Thus, if one writes the change in the normalized internal energy, $u(x,\kappa)/k_\mathrm{B}\T(\kappa)=(1/2)\kappa x^2/k_\mathrm{B}\T(\kappa)$, we find
\begin{equation}
    d\left[\frac{u(x(t),\kappa(t))}{k_\mathrm{B}\T[\kappa(t)]}\right]=\frac{-dq}{k_\mathrm{B}\T[\kappa(t)]}+d\bar{w}.
    \label{Eq:DiffTotalEnergyAppendix}
\end{equation}
Heat is unambiguously defined from the Langevin equation \cite{Sekimoto1998}, consisting of the forces due to dissipative resistance of the host fluid, the thermal fluctuations, and the athermal (active fluctuations), acting along a displacement $dx$,
\begin{equation}
    d q(t) = [\gamma \dot x(t) - \sqrt{2k_\mathrm{B} T \gamma}\xi(t) - \eta(t)]\circ dx(t).
\end{equation}
Upon inserting the R.H.S. of the Langevin equation instead of $\gamma\dot x(t)$, we have simply
\begin{equation}
    d q(t) = -\kappa(t) x(t) \circ dx(t).
    \label{Eq:AppHeat}
\end{equation}
The last term in Eq.~(\ref{Eq:DiffTotalEnergyAppendix})
\begin{equation}
    d\bar w=\frac12\frac d{dt}\left[\frac{\kappa(t)}{k_\mathrm{B}\T[\kappa(t)]}\right]x^2(t)dt
\end{equation}
is a generalized normalized work\,---\,it includes both the stiffness change and the stiffness-induced temperature change.

Since the change in $\kappa(t)$ (and thus $k_\mathrm{B}\T[\kappa(t)]$) is instantaneous, the accumulated work along a transformation from $\kappa(t<0)=\kappa_\mathrm{i}$ to $\kappa(t>0)=\kappa_\mathrm{f}$ is given by Eq.~(\ref{Eq:stepWork}). The second law of thermodynamics is an inequality on the average generalized work. Using the fact that the transformation starts at NESS, $\langle x^2(0)\rangle=k_\mathrm{B}\T(\kappa_\mathrm{i})/\kappa_\mathrm{i}$, we find the average generalized work
\begin{equation}
    \langle\bar w\rangle=\frac12\left[\frac{\kappa_\mathrm{f}}{k_\mathrm{B}\T(\kappa_\mathrm{f})}\frac{k_\mathrm{B}\T(\kappa_\mathrm{i})}{\kappa_\mathrm{i}}-1\right]
    \label{eq:AppWorkStep}
\end{equation}
at the end of the process.
Using the effective Boltzmann factor, Eq.~(\ref{eq:effboltzmann}), we arrive at the change in normalized free energy, $\mathcal{F}(\kappa)/k_\mathrm{B}\T(\kappa)=-\ln[\mathcal{Z}(\kappa)]$
\begin{equation}
    \Delta \left[\frac{F(\kappa(t))}{k_\mathrm{B}\T[\kappa(t)]}\right]=\frac12\ln\left[\frac{\kappa_\mathrm{f}}{k_\mathrm{B}\T(\kappa_\mathrm{f})}\frac{k_\mathrm{B}\T(\kappa_\mathrm{i})}{\kappa_\mathrm{i}}\right].
\end{equation}
Combining, we find the mean total entropy production along a NESS-to-NESS transformation, $\langle\sigma_\mathrm{tot}\rangle=\langle\bar{w}\rangle-\Delta(\mathcal{F}/\T)$
\begin{eqnarray}
    \langle\sigma_\mathrm{tot}\rangle&=&\frac{k_\mathrm{B}}{2}\left\{\frac{\kappa_\mathrm{f}}{k_\mathrm{B}\T(\kappa_\mathrm{f})}\frac{k_\mathrm{B}\T(\kappa_\mathrm{i})}{\kappa_\mathrm{i}}-1\right.\nonumber\\&&\left.-\ln\left[\frac{\kappa_\mathrm{f}}{k_\mathrm{B}\T(\kappa_\mathrm{f})}\frac{k_\mathrm{B}\T(\kappa_\mathrm{i})}{\kappa_\mathrm{i}}\right]\right\}.
\end{eqnarray}
Noting that $x-1-\ln x\geq0$ for $x>0$ (where equality is only satisfied for $x=1$), we confirm that the effective temperature of Eq.~(\ref{Eq:Teff}) satisfies the second law of thermodynamics, $\langle\sigma_\mathrm{tot}\rangle\geq0$. 

\begin{figure}[t!]
    \centering
    \includegraphics[width=1\linewidth]{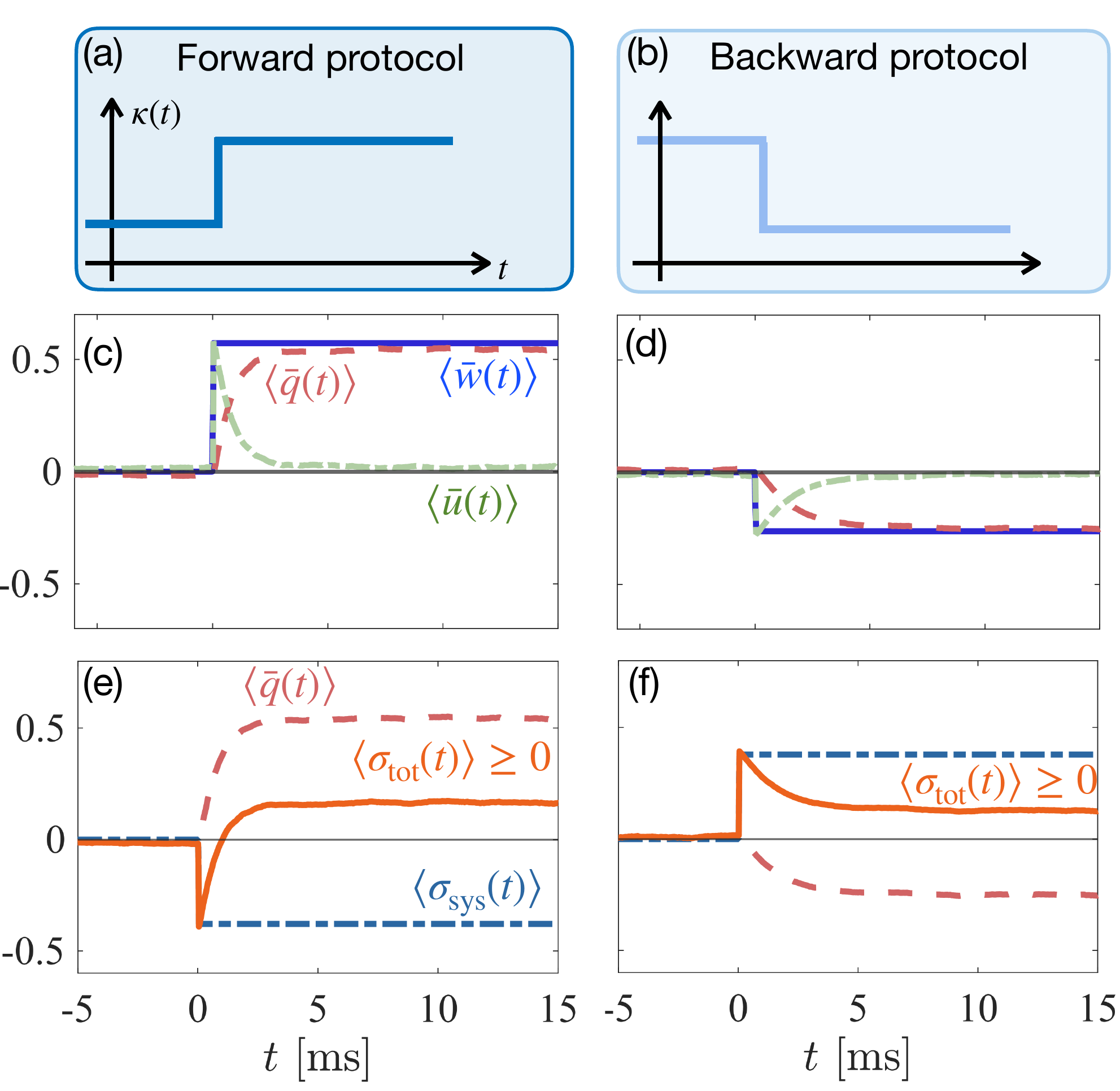}
    \caption{First and second laws on averaged thermodynamic quantities both for (a) the \textit{forward}(compression) and (b) the \textit{backward}(expansion) transformations.
    (c) Ensemble averaged generalized work (blue solid line), heat (red dashed line) and internal energy (light dash-dotted green line) demonstrating the validity if the First Law $\langle \bar u\rangle = \langle \bar w \rangle - \langle \bar q \rangle $ in our system during a forward protocol.
    (d) First Law for the backward protocol
    (e) Ensemble averaged system entropy (blue dash-dotted line), medium entropy (\textit{ie} the heat divided by effective temperature, red dashed line) and total entropy $\sigma_{\rm tot} = \sigma_{\rm sys} + \sigma_{\rm med}$ (orange solid line), demonstrating the second law $\Delta \sigma_{\rm tot} \geq 0$  during a forward protocol.
    (f) second law for a backward protocol.}
    \label{fig:FirstLaw}
\end{figure}

Since the change is abrupt (i.e., irreversible from a classical-thermodynamic perspective), $\langle\sigma_\mathrm{tot}\rangle=0$ only if no change has occurred, $\kappa_\mathrm{i}=\kappa_\mathrm{f}$. However, the effective temperature does capture the quasistatic limit: Namely, if $\Delta\kappa=\kappa_\mathrm{f}-\kappa_\mathrm{i}$ is small, then the extractable work and heat are to leading order linear in small $\Delta\kappa$, $\langle\bar w\rangle,\langle q\rangle\sim\Delta\kappa$, however, the underlying irreversibile cost (dissipated energy) is smaller\,---\,$\langle\sigma_\mathrm{tot}\rangle\sim\Delta\kappa^2$. Thus, were one to perform many small increments in $\Delta\kappa$, they would be able to extract work $\sim1$, while only dissipating energy $\sim\Delta\kappa$.

As shown on Fig.~\ref{fig:FirstLaw}, the mean values of the normalized heat, work and internal energy obey the First Law for both the forward Fig.~\ref{fig:FirstLaw}(a,c) and backward Fig.~\ref{fig:FirstLaw}(b,d) transformations.
The forward compression corresponds to a work $\langle \bar w \rangle = \int \langle d\bar w \rangle $ exerted by the potential on the system.
It is then dissipated as heat $\langle \bar q \rangle = \int \langle \frac{dq}{k_\mathrm{B} \T} \rangle dt$ into the bath.
Conversely, the backward expansion corresponds to a work exerted by the system against the potential, fueled by heat absorbed from the heat bath.
The work extracted during the expansion (Fig.~\ref{fig:FirstLaw}(d)) is smaller than the work injected during the compression (Fig.~\ref{fig:FirstLaw}(c)), which is a consequence of the second law.
More generally, the second law implies the non-negativity of the total entropy at the end of the full process.
It is fulfilled both for the compression Fig.~\ref{fig:FirstLaw}(e) and expansion Fig.~\ref{fig:FirstLaw}(f).
We note that the total entropy can transiently become negative during the transformation, the generalized second law is valid only over the entire process.

\section{Derivation of the fluctuation theorem for work}
\label{App:CFT}

The accumulated stochastic generalized work is written in term of the stochastic position $x(0)$ in Eq.~(\ref{Eq:stepWork}).
This allows to derive the probability distribution of work after a step-like protocol, knowing the distribution of position $p(x)$ at time $t_0$.
This leads to
\begin{equation}
    p(\bar w) = \left([s_{\rm f} - s_{\rm i}] \frac{\bar w}{2}\right)^{-1/2} ~\exp\left[\frac{\mathcal{F}(\kappa_{\rm{i}})}{k_{\rm{B}} \T[\kappa_{\rm{i}}]} - \frac{\bar w}{\frac{s_{\rm f}}{s_{\rm i}} - 1}\right]
\end{equation}
where we use the short notation $s_{\rm i} = \kappa_{\rm i} / k_{\rm B} \T[\kappa_{\rm i}]$, (resp. $s_{\rm f}$).
When performing the time-reversed protocol, $\Delta \kappa$ and $\Delta \T$ simply change sign, which lead to
\begin{equation}
    \bar p(-\bar w) = \left([s_{\rm f} - s_{\rm i}] \frac{\bar w}{2}\right)^{-1/2} ~\exp\left[\frac{\mathcal{F}(\kappa_{\rm f})}{k_{\rm B} \T[\kappa_{\rm f}]} + \frac{\bar w}{\frac{s_{\rm i}}{s_{\rm f}} - 1}\right].
\end{equation}

\begin{figure}[t!]
	\centerline{\includegraphics[width=1\linewidth]{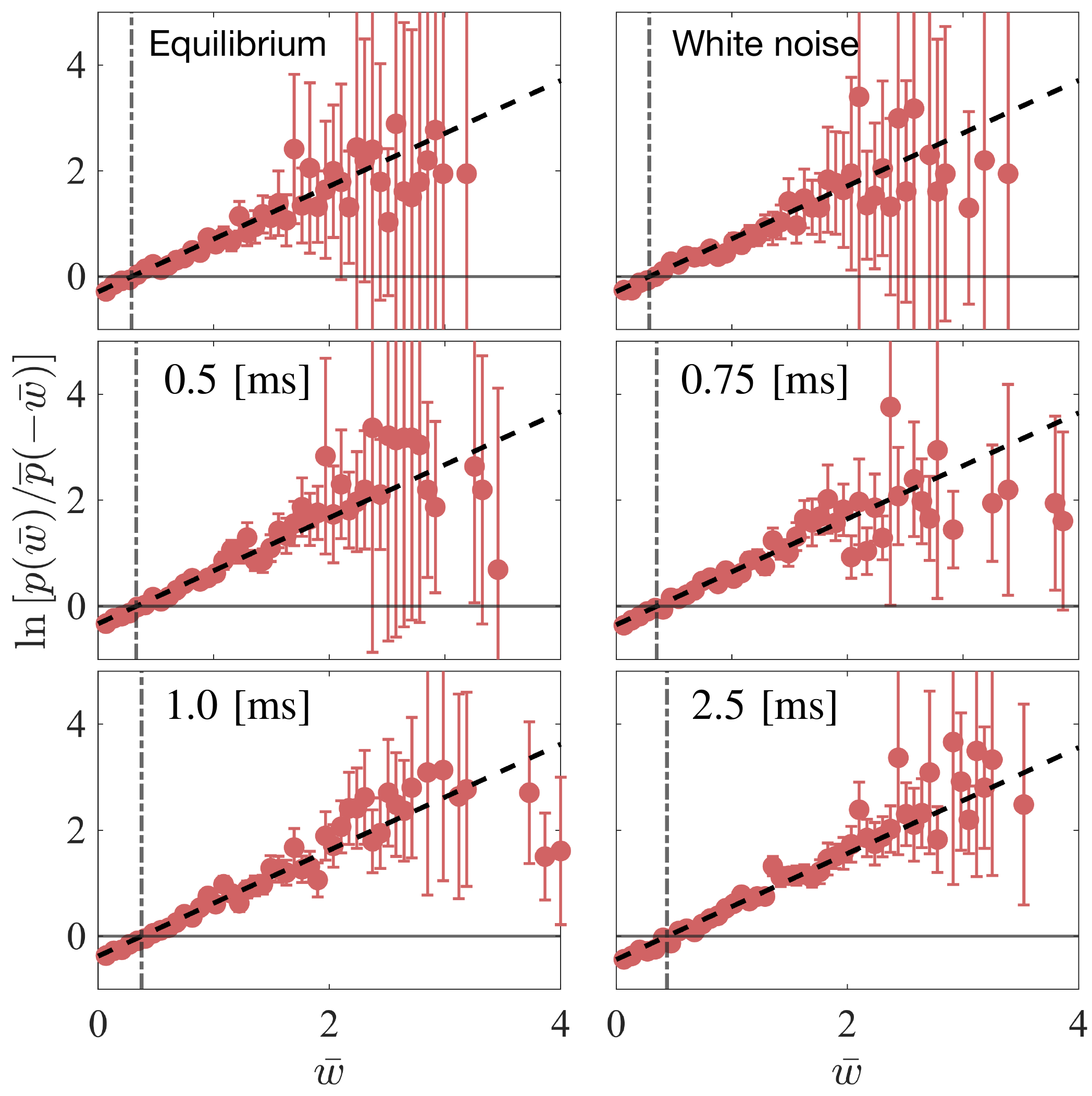}}	
	\caption{
        Fluctuation theorem Eq.~(\ref{Eq:AppFT}) for equilibrium (top left panel) white noise driven (top right panel) and colored noise driven cases, with increasing correlation time (four lower panels).
        The experimentally measured log-ratio (red filled circles) agree within error with the expected result (black dashed line). The normalized free-energy difference (vertical dot-dashed line) agrees with the work $\bar w_0$ for which the log-ratio crosses zero. 
        }
	\label{fig:AllFT}
\end{figure}

Finally, the log-ratio of probabilities of work in the forward and backward process reads
\begin{equation}
    \ln\left[ \frac{p(\bar w)}{\bar p(-\bar w)} \right] = \bar w - \Delta \left( \frac{\mathcal{F}}{k_{\rm B} \T}\right)
    \label{Eq:AppFT}
\end{equation}
where $\Delta \mathcal O \equiv (\mathcal{O}_{\rm f} - \mathcal{O}_{\rm i})$.
This is the detailed FT for our generalized work under a step-like perturbation of stiffness in a nonequilibrium bath.
In Fig.~\ref{fig:CFT}(b) and Fig.~\ref{fig:AllFT}, we shown the experimentally measured ratio $\ln\left[ p(\bar w) / \bar p(-\bar w) \right]$ (symbols) together with the expected result $\bar w - \Delta \left( \mathcal{F} / k_{\rm B} \T\right)$ (lines).
The FT is fulfilled for the various bath correlation times.
The log-ratio intercepts $0$ for a value of work $\bar w_0$ that corresponds to the difference in normalized free-energy difference (vertical dot-dashed lines in Fig.~\ref{fig:AllFT}).

\begin{figure}[t!]
	\centerline{\includegraphics[width=0.9\linewidth]{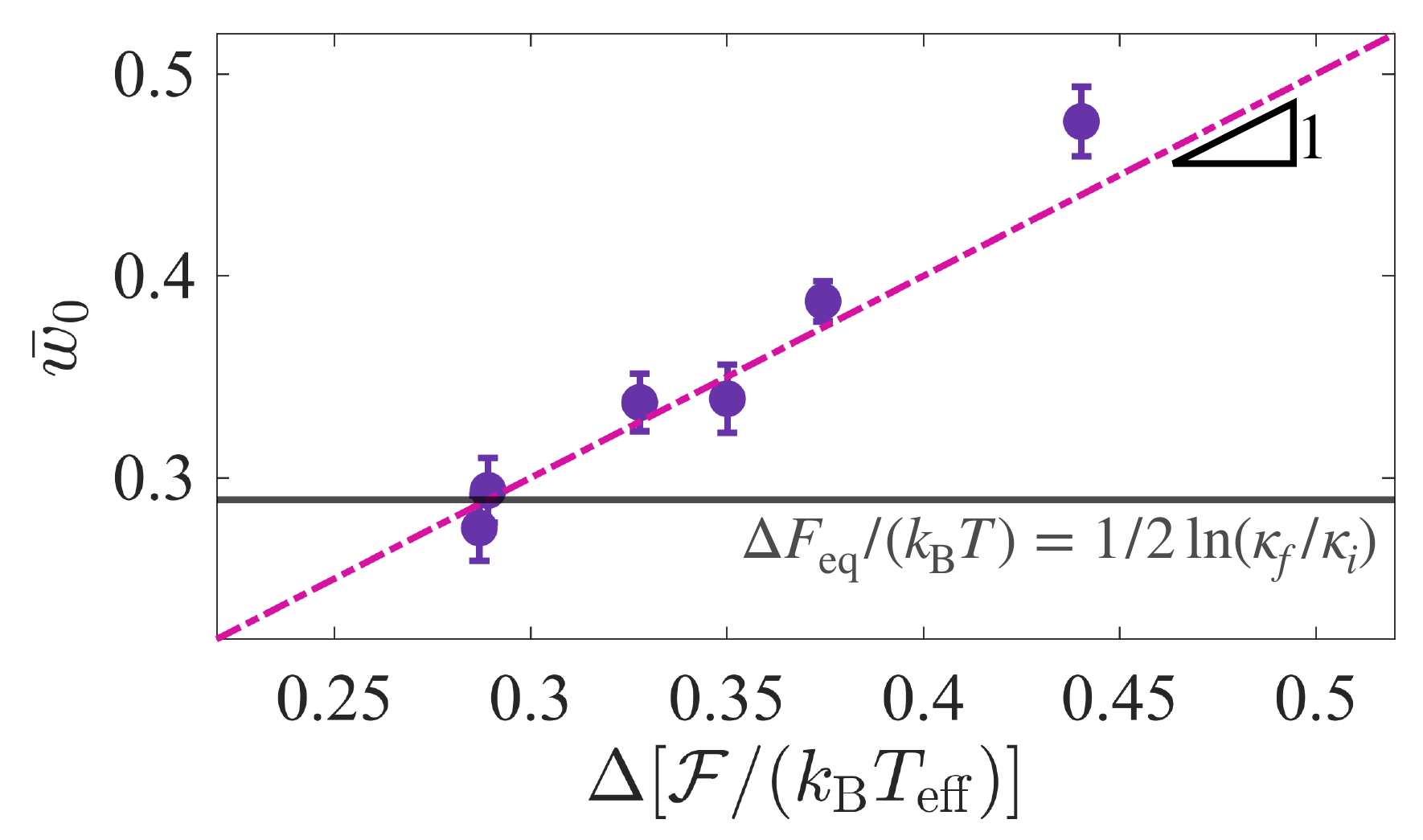}}	
	\caption{Normalized non-equilibrium free-energy estimated from the experimental distributions of work, as the value $\bar w_0$ for which the log-ratio of distributions crosses zero (purple filled circles, it corresponds to the vertical dashed lines in Fig.~\ref{fig:AllFT}) as a function of the exact $\Delta (\mathcal F / k_{\rm B} T_{\rm eff})$. We further underline the equilibrium free-energy difference $\Delta F_{\rm eq} / k_{\rm B} T = 1/2 \ln(\kappa_{\rm f}/\kappa_{\rm i})$ (horizontal solid line) and the expected one-to-one curve for the nonequilibrium cases (pink dot-dashed line).
		}
	\label{fig:DeltaF}
\end{figure}

In Fig.~\ref{fig:DeltaF} we compare quantitatively the values of $\bar w_0$ measured from a linear fit of the experimental log-ratio, to the expected value of the normalized free-energy difference.
The result agree within fitting errors for most noise correlations.
It demonstrates that this FT can serve to measure the normalized free-energy difference (\textit{i.e.} the minimal generalized work exchanged during in the quasistatic transformation) from a finite-time experiment.

\section{Derivation of the variance FRR}
\label{App:VarFRR}

Fluctuation-Response Relations (FRRs) relate the mean response of a system to an external perturbation to the autocorrelation function with the unperturbed destination system.
In its usual form, it considers a perturbation by a constant linear force $f$ and its conjugated variable, the mean position of a particle, $\langle x(t)\rangle$.
Here we take a different approach and focus on a perturbation consisting of a stiffness change and its conjugated variable, the system's second moment $\langle x^2(t) \rangle$.

We focus on the transition between two equilibrium states in the absence of an external drive ($\eta = 0$). We first evaluate the response function $\mathcal{R}(t)=\langle x^2(t) \rangle-\langle x^2 \rangle_{\rm f}$ where $\langle x^2 \rangle_{\rm f} = \langle x^2(t = \infty) \rangle$ is the unperturbed second moment, reached at the end of the relaxation. We multiply both sides of Eq.~(\ref{Eq:Langevin}) by $x(t)$ with a Stratonovich product, and take the ensemble average. This results in the equation $(1/2)d\langle x^2(t)\rangle=-\gamma^{-1}\kappa_\mathrm{f}\langle x^2(t)\rangle+D$~\cite{book:schuss}. Following equipartition, the initial and final conditions read $\kappa_\mathrm{i}\langle x^2(0)\rangle=\kappa_\mathrm{f}\langle x^2(\infty)\rangle=k_\mathrm{B}T$.
Solving the equation for the variance, we find
\begin{equation}
    \langle x^2(t) \rangle = \left( \frac{k_\mathrm{B} T}{\kappa_\mathrm{i}} - \frac{k_\mathrm{B} T}{\kappa_\mathrm{f}} \right) e^{-2\kappa_\mathrm{f}t/\gamma} + \frac{k_\mathrm{B} T}{\kappa_\mathrm{f}}.
\end{equation}
The autocorrelation function $C_{xx}(t)=\langle x(t)x(0)\rangle$ is computed within equilibrium under the final stiffness $\kappa_\mathrm{f}$, therefore, $\kappa_\mathrm{f}\langle x^2(0)\rangle=k_\mathrm{B}T$ for these purposes. This time, we multiply \eqref{Eq:Langevin} by $x(0)$ (which is independent of $\xi(t)$ for all $t>0$), and take the ensemble average. This yields $(d/dt)\langle x(t)x(0)\rangle=-\gamma^{-1}\kappa_\mathrm{f}\langle x(t)x(0)\rangle$. Solving for the autocorrelation, we find
\begin{equation}
    \langle x(t)x(0)\rangle = \frac{k_\mathrm{B} T}{\kappa_\mathrm{f}}  e^{-\kappa_\mathrm{f} t/\gamma}.
\end{equation}
Upon comparing the exponential decays of both expressions, we find
\begin{equation}
    \mathcal{R}(t) = \frac{\kappa_\mathrm{f} (\kappa_\mathrm{f} - \kappa_\mathrm{i})}{k_\mathrm{B} T \kappa_\mathrm{i}} C_{xx, \rm{eq}}^2(t)  = \frac{\Delta \kappa}{k_\mathrm{B}T}\left(1 + \frac{\Delta \kappa}{\kappa_\mathrm{i}}\right) C_{xx, \rm{eq}}^2(t)
    \label{eq:FRR_app}
\end{equation}
Eq.~(\ref{eq:FRR_app}) is an FRR suited to study stiffness changes.
It relates the response of the system to a perturbation to the unperturbed correlation function of the final system.

\begin{figure}[t!]
    \centering
    \includegraphics[width=1\linewidth]{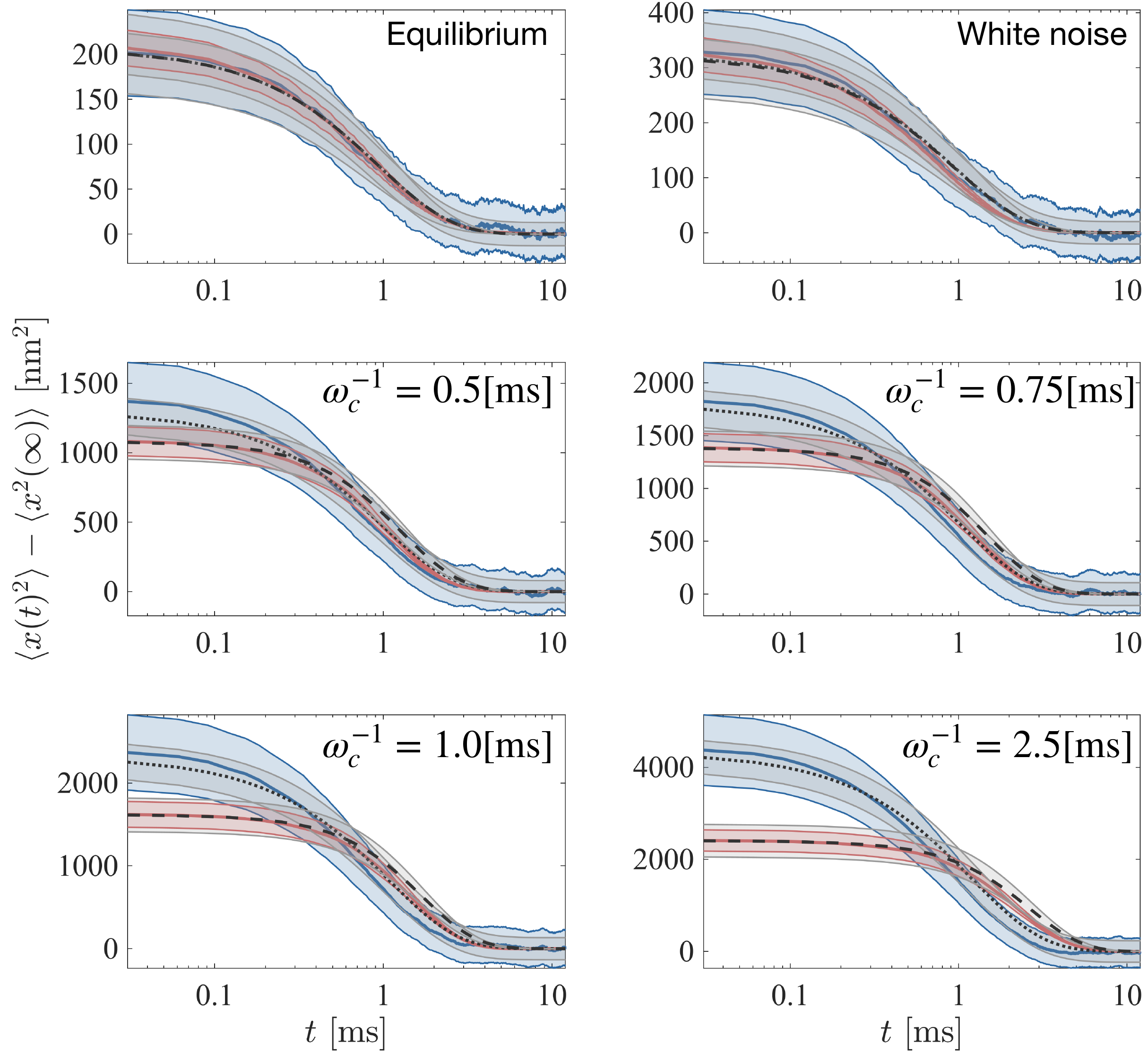}
    \caption{Test of the FRR Eq.~(\ref{eq:FRR_app}), normalized by effective temperature. The measured evolution of the response function $\mathcal R(t)$ (blue solid line, with calibration errors and statistical errors measured using a $\chi^2$-test with a $3\sigma$ confidence interval, as a blue shaded region) shown together with its analytical expression (the time-dependent second-moment is given by Eq.(\ref{eq:NeqVar}) in Appendix) (black dotted line, with propagated error on the fitted $\kappa_\mathrm{i}$ and $\kappa_\mathrm{f}$ shown as a dark shaded area). We also plot the measured rescaled squared correlation function $[\kappa_\mathrm{f} (\kappa_\mathrm{f} - \kappa_\mathrm{i})]/[k_\mathrm{B} \T(\kappa_\mathrm{f}) \kappa_\mathrm{i}] C_{xx,}^2(t)$ (red solid line, with calibration error and the errors on the physical parameters as a red shaded area) together with the analytical result (the time-dependent correlation function is given by Eq.(\ref{eq:NeqCxx}) in Appendix) (black dashed line, with propagated error on the fitted $\kappa_\mathrm{i}$ and $\kappa_\mathrm{f}$ shown as a dark shaded area).
    The upper left panel corresponds to the case of thermal equilibrium $\eta = 0$, the upper right panel corresponds to a white noise driving, and all other cases correspond to correlated noise drivings of increasing correlation time $[0.5, ~0.8, ~1, ~2.5]~\rm{ms}$.
    }
    \label{fig:AllFRR}
\end{figure}

\begin{figure}[t!]
    \centering
    \includegraphics[width=1\linewidth]{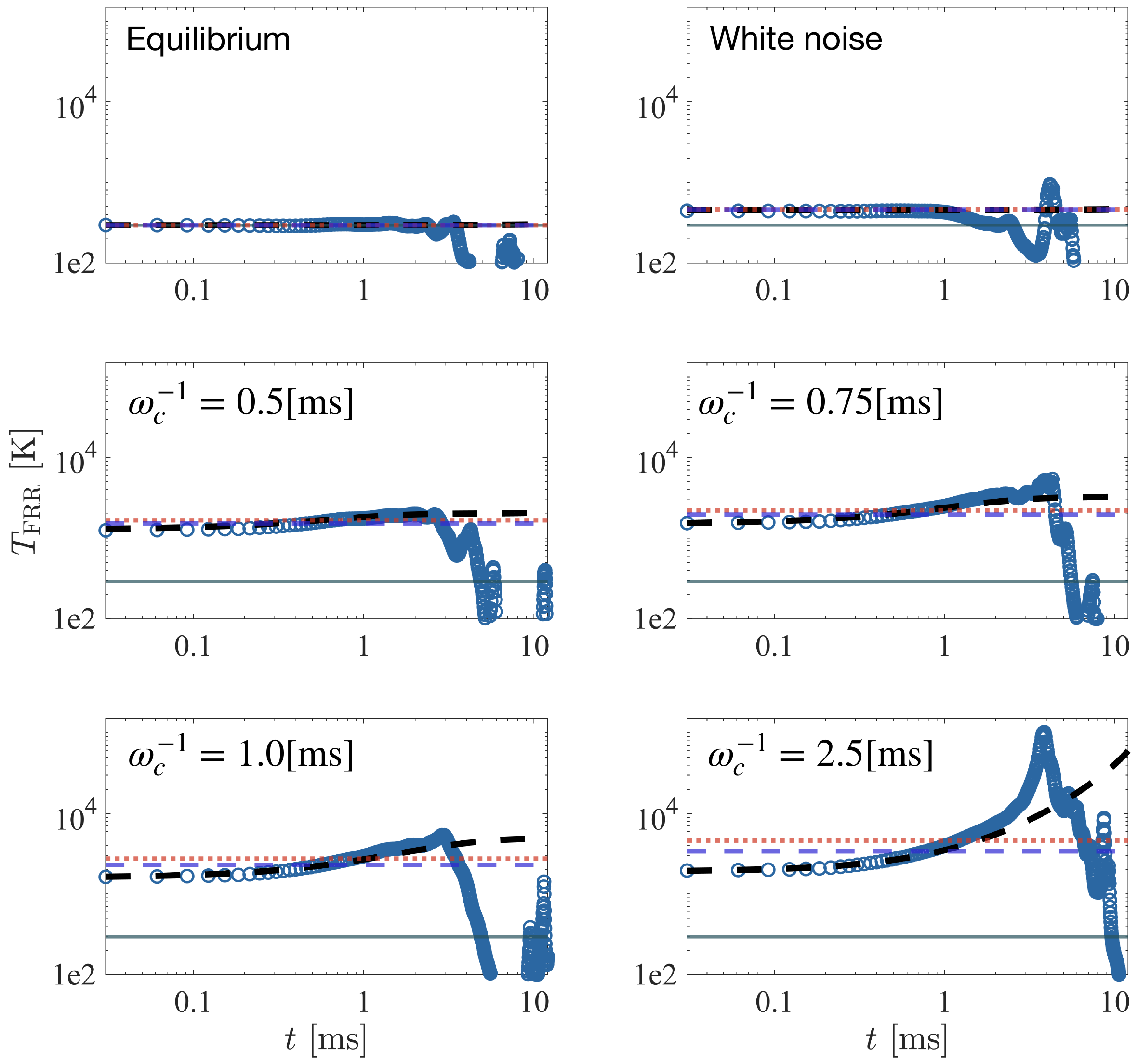}
    \caption{Temperature measured from the FRR following Eq.~(\ref{Eq:T_FRR}) on experimental data (blue circles) and using analytical expressions for $\mathcal{R}$ and $C_{xx}$ (Eqs.~(\ref{eq:NeqVar}) and (\ref{eq:NeqCxx})) for the response and correlation functions (dashed black line). We observe good agreement until the long-time limit, where the ratio of two quantities near zero produces a noisy output.
    Room temperature $T$ is underlined in each graph (horizontal gray-blue solid line), along with $\T(\kappa_\mathrm{i})$ (red dotted line) and $\T(\kappa_\mathrm{f})$ (blue dashed line).
    }
    \label{fig:AllFRRTemp}
\end{figure}

We propose to use the validity of the equality Eq.~(\ref{eq:FRR_app}) to distinguish effective equilibrium (memory-less white-noise driving) from inherent nonequilibrium. Namely, for true equilibrium, the response and correlations will not only share the value at $t=0$ but will also have the same time-dependence for all $t>0$.
On the other hand, we expect Eq.~(\ref{eq:FRR_app}) to break when additional external noises drive the system in a nonequilibrium state.
This FRR-breaking changes nature, depending on whether the system is Markovian or not.
In Markovian nonequilibrium cases (white noise driving), Eq.~(\ref{eq:FRR_app}) may be retrieved by modifying the prefactor of the correlation function by introducing an effective temperature.
In non-Markovian nonequilibrium cases (noise with memory, but instantaneous viscosity kernel \cite{marconi2017heat}), the variance and correlation will have distinct time dependence.
Then, the FRR cannot be recovered, unless a time-dependent effective temperature (along the relaxation of the system, as seen Fig.~\ref{fig:AllFRRTemp}, in variance with our state-dependent temperature, which is constant during the relaxation of the system) is introduced, as a signature of the nonequilibrium and non-Markovian nature of the process \cite{dieterich2015single}.

In Fig.~\ref{fig:AllFRR}, we show the outcomes of this FRR test for all cases studied in this paper. 
Here, $\T(\kappa_\mathrm{f})$ is the effective temperature in the final steady state inside the prefactor of the correlation function. Naturally, for $\eta =0$, it coincides with room temperature $T$.
We experimentally verified that at thermal equilibrium (upper left panel), both variance and squared correlation coincide along the entire probed dynamical range.
They furthermore fully agree with the analytical results plotted using the physical parameters extracted from the experiment.
In the case of an additional artificial white noise (upper right panel), using the equipartition-based effective temperature in the prefactor of the correlation function suffices to retrieve the FRR.
This shows that, if the system is driven out of equilibrium by the external noise, it effectively cannot be distinguished from an equilibrium system at a higher temperature.
It can also be seen as a consequence from the fact that a white-noise-driven system is out of equilibrium with respect to the thermal bath but stays Markovian.
In the case of correlated noise, the FRR is broken. The disagreement worsens with increased correlation time $\omega_\mathrm{c}^{-1}$.
Using the effective temperature is not enough to make both sides of Eq.~(\ref{eq:FRR_app}) coincide, revealing the intrinsically non-Markovian nature of the dynamics at play.\\

\textit{FRR-based effective temperature}-
We introduce an alternative definition for an effective temperature, unrelated to the one of Eq.~(\ref{Eq:Teff}). This is motivated by frequency-dependent temperatures used in the context of glassy systems \cite{cugliandolo2011effective, crisanti2003violation, joubaud2009aging}, granular gases with mixed time-scales \cite{gnoli2014nonequilibrium}, but also active matter \cite{zamponi2005generalized, SzamelPRE14} (including also the AOUP model considered here \cite{szamel2023single}).
This definition is directly based on the FRR, defining the temperature as the ratio among the correlation and response functions~\cite{cugliandolo2011effective}, thus enforcing the FRR.

We rearrange the FRR, Eq.~(\ref{Eq:FRR}), leading to the definition
\begin{equation}
    T_{\rm FRR}(t) \equiv \frac{\kappa_\mathrm{f}(\kappa_\mathrm{f} - \kappa_\mathrm{i}) C_{xx}^2(t)}{k_\mathrm{B} \kappa_\mathrm{i} \mathcal{R}(t)}.
    \label{Eq:T_FRR}
\end{equation}

In Markovian systems, this temperature coincides with the equipartition temperature, Eq. ~(\ref{Eq:Teff}), by construction. Otherwise, upon computing $C_{xx}(t)$ and $\mathcal{R}(t)$, one may expect complicated time-dependencies, and even more so simultaneous $\kappa_\mathrm{i}$ and $\kappa_\mathrm{f}$ dependence, thus implying history dependence.

In Fig.~\ref{fig:AllFRRTemp}, we show $T_{\rm FRR}(t)$ as a function of time for the same cases as in Fig.~\ref{fig:AllFRR}.
We superimpose the experimental and analytical values of this ratio, as well as room temperature and the two values of the state-dependent temperature used in this work.
In the case of thermal equilibrium (Fig.~\ref{fig:AllFRRTemp}, top left panel) and white-noise driving (Fig.~\ref{fig:AllFRRTemp} top right panel), all definitions of temperature (except the room temperature in the latter case) coincide, as expected in these (pseudo-) equilibrium scenarios.
In the case of correlated noise, $T_{\rm FRR}$ is a function of time, which does not coincide with either state-dependent temperature $\T(\kappa)$ defined here, even at the origin.
At late-time, the ratio of two vanishing quantities produces large experimental errors, as visible on the figure.

\section{Derivation of the autocorrelation and response function}

The system under study can be modeled by the following uni-directionally coupled stochastic differential equations (SDEs) Eqs.(\ref{Eq:Langevin}) and (\ref{Eq:Noise}).
The choice of normalization of the external force $\eta(t)$ ensures a constant variance, even when the correlation time $\omega_c^{-1}$ is modified.
As discussed in Appendix \ref{App:Exp}, the positional PSD
\begin{equation}
    S_{x x}[\omega] = \frac{1}{(\omega_0^2+\omega^2)}\left( 2D + 
    \frac{2\sigma_\eta^2\omega_c}{\gamma^2(\omega_c^2+\omega^2)} \right)
\end{equation}
where the first term corresponds to the thermal equilibrium Lorenzian profile, with integral $D/\omega_0$, while the second term introduces the correction due to the presence of the correlated forcing $\eta(t)$.

\begin{widetext}

Via Wiener–Khinchin theorem, the Fourier transform of $S_{xx}[\omega]$ corresponds to the autocorrelation function \cite{Goerlich2022}
\begin{equation}
    C_{xx}(t) = \frac{D}{\omega_0}  e^{-\omega_0 t} + \frac{\sigma_\eta^2  \omega_c}{\gamma^2 \omega_0 (\omega_c^2 - \omega_0^2)} \left( e^{-\omega_0 t} - \frac{\omega_0}{\omega_c} e^{-\omega_c t} \right)
    \label{eq:NeqCxx}
\end{equation}
where the first term $C_{xx, \rm{eq}}(t) = \frac{D}{\omega_0}  e^{-\omega_0 t}$ corresponds to the equilibrium correlation function.

We now consider the case of a step-like change of stiffness, with the characteristic inverse time $\omega_0$ changing instantaneously from $\omega_{\rm i}$ to $\omega_{\rm f}$ at $t = 0$.
The solution of Eq.~(\ref{Eq:Langevin}) for $t > 0$ reads
\begin{equation}
    x(t) = x(0) e^{-\omega_{\rm f} t } + \int_0^t dt' e^{-\omega_{\rm f} (t - t')} \left( \sqrt{2 D} \xi(t') + \frac{1}{\gamma}\eta(t') \right).
\end{equation}
We compute the variance $\langle x^2(t) \rangle$ by taking the square of the solution.
\begin{align*}
    \langle x^2(t) \rangle = &\langle x(0)^2\rangle e^{-2\omega_{\rm f} t} + \frac{2}{\gamma} \int_0^t dt_1 e^{–\omega_{\rm f} (2t - t_1)} \langle x(0) \eta(t_1)\rangle\\
    &+ \int_0^t dt_1 \int_0^t dt_2 e^{-\omega_{\rm f}(2t - t_1 - t_2)} \left( \delta(t-t_1) + \frac{1}{\gamma^2} C_{\eta\eta}(t_1 - t_1)\right)
\end{align*}
where $ C_{\eta\eta}(t_1 - t_1) \equiv \langle \eta(t_1)\eta(t_2)\rangle$. In order to compute the correlation between $x(0)$ and $\eta(t_1>0)$ we use the solution $x(t<0)$ before the change of stiffness
\begin{equation}
    x(t<0) = \int_{-\infty}^t dt_2 e^{-\omega_{\rm i} (t - t_2)} \left( \sqrt{2 D} \xi(t_2) + \frac{1}{\gamma}\eta(t_2) \right).
\end{equation}
which allows to obtain $ \langle x(0) \eta(t_1)\rangle = \frac{\sigma_\eta^2}{\gamma^2 (\omega_{\rm i} + \omega_c)}e^{-\omega_c t_1}$ which can be inserted into the equation for the variance, leading to

\begin{equation}
  \label{eq:NeqVar}
  \begin{gathered}[b]
    \langle x^2(t) \rangle = \left( \frac{D}{\omega_{\rm i}} + \frac{\sigma_\eta^2}{\gamma^2\omega_{\rm i}(\omega_{\rm i} + \omega_c)} \right) e^{-2 \omega_{\rm f} t} + \frac{D}{\omega_{\rm f}}\left( 1 - e^{-2\omega_{\rm f} t}\right) \\
    + \frac{\sigma_\eta^2}{\gamma^2} \left( \frac{1}{\omega_{\rm f}(\omega_{\rm f}+\omega_c)} + \frac{2 e^{-2\omega_{\rm f} t}}{\omega_{\rm f}(\omega_{\rm f} - \omega_c)} - \frac{2e^{-(\omega_{\rm f} + \omega_c)t}}{\omega_{\rm f}^2 - \omega_c^2} \right)\\
    + \frac{ 2\sigma^2}{\gamma^2(\omega_{\rm i} + \omega_c)(\omega_{\rm f} - \omega_c)} \left( e^{-(\omega_{\rm f} + \omega_c)t} - e^{-2\omega_{\rm f} t} \right)
  \end{gathered}
\end{equation}

\end{widetext}
For $t=0$ we obtain 
\begin{equation}
    \langle x(0) \rangle = \frac{D}{\omega_{\rm i}} + \frac{\sigma_\eta^2}{\gamma^2 \omega_{\rm i} (\omega_{\rm i} + \omega_c)}
\end{equation}
and in the limit of $t\rightarrow \infty$ we obtain as expected
\begin{equation}
    \langle x(t \rightarrow+\infty) \rangle = \frac{D}{\omega_{\rm f}} + \frac{\sigma_\eta^2}{\gamma^2 \omega_{\rm f} (\omega_{\rm f} + \omega_c)}.
\end{equation}
In the case where $\sigma_\eta^2 = 0$ we retrieve the single exponential decay of a transient between states at thermal equilibrium $\langle x^2(t) \rangle_{\rm eq} = \left( \frac{D}{\omega_{\rm i}} - \frac{D}{\omega_{\rm f}} \right) e^{-2\omega_{\rm f}t} + \frac{D}{\omega_{\rm f}}$.

\section{Derivation of the generated entropy for $\kappa(t)$ in a non-equilibrium bath}

As we introduced in the main text, along a protocol of changing stiffness $\kappa(t)$, both $\kappa$ and $\T(\kappa)$ will evolve in time.
At every instant of time, it is instructive to consider the instantaneous NESS distribution $p(x,\kappa(t))$, corresponding to the distribution which will be reached if the system is allowed to relax under a potential of fixed stiffness $\kappa(t)$, imposing a temperature $\T[\kappa(t)]$.

The stochastic system entropy is related to the instantaneous NESS distribution evaluated at the stochastic position is given by Eq.~(\ref{eq:SysEntropy}).
It depends on time via $\kappa(t)$, $\T[\kappa(t)]$ and $x(t)$. Its differential increment reads, using the stochastic chain rule~\cite{book:schuss},
\begin{align}
    d\sigma_{\rm sys}(t) = &\left( 1 - \frac{\kappa(t) x^2(t)}{k_\mathrm{B} \T[\kappa(t)]}\right) \frac{k_\mathrm{B} d\T[\kappa(t)]}{2 \T[\kappa(t)]}\nonumber\\
    &- \left( 1 - \frac{\kappa(t) x^2(t)}{k_\mathrm{B} \T[\kappa(t)]}\right) \frac{k_\mathrm{B} d\kappa(t)}{2\kappa(t)}\nonumber\\
    &+ \frac{\kappa(t) x(t)}{\T[\kappa(t)]}  \circ dx(t).
    \label{eq:App_dSysEntropy}
\end{align}
The last term can be identified as the medium entropy, $dq(t)/\T[\kappa(t)]$, as follows. The heat is unambiguously defined from the Langevin equation \cite{Sekimoto1998}, as Eq.~(\ref{Eq:AppHeat}).

Equipped with Eq.~(\ref{eq:App_dSysEntropy}), we compute the entropy production, $d\sigma_{\rm tot} = d\sigma_{\rm sys} + dq/\T$,
\begin{equation*}
    d\sigma_{\rm tot}(t) = \left( 1 - \frac{\kappa(t) x^2(t)}{k_\mathrm{B} \T[\kappa(t)]}\right) \left( \frac{k_\mathrm{B} d\T[\kappa(t)]}{2 \T[\kappa(t)]} - \frac{k_\mathrm{B} d\kappa}{2\kappa(t)}\right).
\end{equation*}
As a functional of $x(t)$, it is a stochastic quantity.
Its ensemble average value $dS_{\rm tot} \equiv \langle d \sigma_{\rm tot} \rangle$ is computed by replacing $x^2(t)$ by the time-dependent variance.
It vanishes only if $\big(1 - \kappa(t) \langle x^2(t)\rangle/k_\mathrm{B} \T[\kappa(t)]\big)$ is zero (\textit{i.e.} when generalized equipartition is fulfilled), which is a measure of deviation to quasistaticity.
As such, it does correspond to a sensible measure of irreversibility in this effective equilibrium framework.

The system entropy Eq.~({\ref{eq:SysEntropy}}) can be directly related to thermodynamic quantities as
\begin{equation}
    \sigma_{\rm sys}(t) = \frac{u(x, t) - \mathcal{F}(t)}{\T[\kappa(t)]}
\end{equation}
where the instantaneous stochastic internal energy of the system reads $u(x, t) = (1/2)\kappa(t) x^2(t)$ and $\mathcal{F}$ is the non-equilibrium free energy.
Using this definition, Eq.~(\ref{eq:App_dSysEntropy}) can be recast as
\begin{align*}
    d\sigma_{\rm sys} &= d \left( \frac{u}{\T} \right) - d \left( \frac{\mathcal{F}}{\T} \right)\\
      &= \frac{x^2(t)}{2 k_\mathrm{B} \T[\kappa(t)]} d\kappa - \frac{\kappa(t) x^2(t)}{2k_\mathrm{B}\T^2[\kappa(t)]}d\T[\kappa(t)] \\
      & \;\;\;\;\; - \frac{dq}{\T[\kappa(t)]} - d \left( \frac{\mathcal{F}}{\T} \right).
      \label{eq:AppSysEntropy2}
\end{align*}

This clearly underlines a key thermodynamic quantity, which we interpret as a generalized (dimensionless $\bar w=w/k_\mathrm{B}\T$) non-equilibrium work, the differential form of which
\begin{equation}
    d\bar w = \frac{x^2(t)}{2 k_\mathrm{B} \T[\kappa(t)]} d\kappa - \frac{\kappa(t)x^2(t)}{2k_\mathrm{B}\T^2[\kappa(t)]}d\T[\kappa(t)],
    \label{eq:AppWork}
\end{equation}
contains the change internal energy due to the variation of the external parameter $\kappa$ and the associated change in $\T[\kappa(t)]$.
The first term in the work is the usual mechanical work involved in changing the stiffness of a harmonically trapped Brownian object \cite{Sekimoto1998} while the second term corresponds to the thermal or entropic work, introduced in \cite{Rademacher2022}.
This definition allows the usual relation between total entropy production and dissipated work Eq.~(\ref{Eq:totalEntropy}).
This dissipated non-equilibrium work is only zero in the reversible limit, when the work engaged in the transformation exactly reaches the non-equilibrium free energy difference associated with this transformation.

\end{document}